\documentclass[aps,amsfonts,reprint,tightenlines,amssymb,superscriptaddress,twocolumn, prapplied]{revtex4-2}  

\usepackage{graphicx}% Include figure files
\usepackage{dcolumn}% Align table columns on decimal point
\usepackage{bm}% bold math
\usepackage{mathtools}
\usepackage{stmaryrd}

\usepackage{qcircuit}
\usepackage[caption=false]{subfig}

\usepackage{algorithm}
\usepackage[noend]{algpseudocode}
\algrenewcommand\algorithmicdo{}

\makeatletter
\renewcommand{\ALG@name}{Procedure}
\makeatother

\usepackage[linktocpage=true,
  colorlinks=true, 
  pdfborder={0 0 0},
  linkcolor=blue,
  citecolor=red,
  filecolor=yellow,
  urlcolor=blue,
  bookmarks,
  pdfauthor={},
]{hyperref}

\usepackage{orcidlink}
\usepackage{xcolor}
\usepackage{pagecolor}
\usepackage{tabularx}
\usepackage{colortbl} 
\usepackage{xcolor}
\usepackage{ifthen}
\newcounter{is_qcircuit_used}
\begin{document}

\preprint{APS/123-QED}

\title{Encoded probabilistic imaginary-time evolution on a trapped-ion quantum computer for ground and excited states of spin qubits}

\author{Hirofumi Nishi\orcidlink{0000-0001-5155-6605}}
\email{hnishi@quemix.com}
\affiliation{
Quemix Inc.,
Taiyo Life Nihombashi Building,
2-11-2,
Nihombashi Chuo-ku, 
Tokyo 103-0027,
Japan
}
\affiliation{
Department of Physics,
The University of Tokyo,
Tokyo 113-0033,
Japan
}

\author{Yuki Takei}
\affiliation{Platform Laboratory for Science \& Technology, Asahi Kasei Corporation,
2-1, Samejima, Fuji, Shizuoka, 416-8501, Japan}
\affiliation{Informatics Initiative, Asahi Kasei Corporation,
3-1-1 Shibaura, Minato-ku, Tokyo, 108-0023, Japan
}

\author{Taichi Kosugi\orcidlink{0000-0003-3379-3361}}
\affiliation{
Quemix Inc.,
Taiyo Life Nihombashi Building,
2-11-2,
Nihombashi Chuo-ku, 
Tokyo 103-0027,
Japan
}
\affiliation{
Department of Physics,
The University of Tokyo,
Tokyo 113-0033,
Japan
}

\author{Shunsuke Mieda}
\affiliation{Platform Laboratory for Science \& Technology, Asahi Kasei Corporation,
2-1, Samejima, Fuji, Shizuoka, 416-8501, Japan}
\affiliation{Informatics Initiative, Asahi Kasei Corporation,
3-1-1 Shibaura, Minato-ku, Tokyo, 108-0023, Japan
}

\author{Yutaka Natsume}
\affiliation{Informatics Initiative, Asahi Kasei Corporation,
3-1-1 Shibaura, Minato-ku, Tokyo, 108-0023, Japan
}
\affiliation{Platform Laboratory for Science \& Technology, Asahi Kasei Corporation,
2-1, Samejima, Fuji, Shizuoka, 416-8501, Japan}

\author{Takeshi Aoyagi\orcidlink{0000-0001-9229-4226}}
\affiliation{Informatics Initiative, Asahi Kasei Corporation,
3-1-1 Shibaura, Minato-ku, Tokyo, 108-0023, Japan
}

\author{Yu-ichiro Matsushita\orcidlink{0000-0002-9254-5918}}
\affiliation{
Department of Physics,
The University of Tokyo,
Tokyo 113-0033,
Japan
}
\affiliation{
Quemix Inc.,
Taiyo Life Nihombashi Building,
2-11-2,
Nihombashi Chuo-ku, 
Tokyo 103-0027,
Japan
}
\affiliation{Quantum Materials and Applications Research Center,
National Institutes for Quantum Science and Technology (QST),
2-12-1 Ookayama, Meguro-ku, Tokyo 152-8550, Japan
}
\affiliation{
Laboratory for Materials and Structures,
Institute of Innovative Research,
Tokyo Institute of Technology,
Yokohama 226-8503,
Japan
}

\date{\today}

\begin{abstract}
In this study, we employed a quantum computer to solve a low-energy effective Hamiltonian for spin defects in diamond (so-called NV centre) and wurtzite-type aluminium nitride, which are anticipated to be qubits. 
The probabilistic imaginary-time evolution (PITE) method, designed for use in a fault-tolerant quantum computer (FTQC) era, was employed to calculate the ground and excited states of the spin singlet state, as represented by the effective Hamiltonian.  It is difficult to compute the spin singlet state correctly using density functional theory (DFT), which should be described by multiple Slater determinants.
To mitigate the effects of quantum errors inherent in current quantum computers, we implemented a $\llbracket n+2,n,2 \rrbracket$ quantum error detection (QED) code called the Iceberg code.
Despite the inevitable destruction of the encoded state resulting from the measurement of the ancilla qubit at each PITE step, we were able to successfully re-encode and recover the logical success state.
In the implementation of the PITE, it was observed that the effective Hamiltonian comprises large components of the diagonal part and a relatively small non-diagonal part, which is frequently the case with quantum chemistry calculations. An efficient implementation of Hamiltonian simulations, in which the diagonal components dominate, was developed on a quantum computer based on the second-order Trotter-Suzuki decomposition. This is the first instance of an encoded PITE circuit being executed on a trapped-ion quantum computer. Our results demonstrate that QED effectively reduces quantum errors and that we successfully obtained both the ground and excited states of the spin singlet state. Our demonstration clearly manifests that Zr$_{\rm Al}$V$_{\rm N}$, Ti$_{\rm Al}$V$_{\rm N}$, and Hf$_{\rm Al}$V$_{\rm N}$ defects have a high potential as spin qubits for quantum sensors.
\end{abstract}

\maketitle 

\section{Introduction}
The field of quantum technology has made noteworthy advancements in recent years, with notable developments observed in areas including computing, sensing, and communications. 
In particular, the application of quantum computers in material calculations offers the promise of performing highly accurate simulations of molecular and material properties, which are crucial for understanding and designing new materials.
Hamiltonian simulation \cite{Seth1999Science, Abrams1997PRL, Kassal2008PNAS, Childs2012QIC, Berry2015PRL, Low2019Quantum, Childs2021PRX} enables the modelling of a system's quantum behaviour, thereby facilitating insights into its dynamics and interactions. 
Quantum phase estimation (QPE) is indispensable for determining the eigenvalues of Hamiltonians, which are directly related to the ground state energy of the system \cite{Kitaev1995arXiv, Abrams1999PRL, Ding2023Quantum, Ding2023PRXQ}. 
Efficient state preparation is a prerequisite for QPE, and a number of approaches have been proposed, including adiabatic time evolution \cite{Nishimori1998PRE, Farhi2000arXiv, AspuruGuzik2005Science, Nishiya2024PRA}, variational quantum eigensolver (VQE) \cite{Peruzzo2014Ncom, Farhi2014arXiv}, quantum eigenvalue transformation of unitary matrices (QETU) \cite{Dong2022PRXQ}, and imaginary-time evolution (ITE) on quantum computers \cite{gingrich2004non, terashima2005nonunitary, Mcardle2019npjQI, Motta2020NPhys, Benedetti2021PRR, Kosugi2022PRR, Leadbeater2024QST, Xie2024PRA}.
These algorithms can be used to calculate a variety of physical quantities as key components \cite{Kosugi2020PRA, Kosugi2020PRRes, Sun2024npjQI, Sakuma2024PRA}.

In recent years, there has been a substantial increase in the number of qubits for various physical platforms, including superconductivity \cite{Nakamura1999Nature, Arute2019Nature, Ming2021Science}, trapped ions \cite{Moses2023PRX, DeCross2024arXiv}, and neutral atoms \cite{Chew2022NatPhot, Bluvstein2024Nature, Manetsch2024arXiv}. 
As a result of the limitations of noisy intermediate-scale quantum (NISQ) devices and the increasing number of qubits, there has been a growing interest in demonstrating error-corrected logical qubits and logical gate operations \cite{Bluvstein2024Nature, Self2024NPhys, Gupta2024Nature, Konno2024Science}. These developments are pushing the boundaries of what is possible with current technology. 
Furthermore, quantum computer hardware can now be accessed via the cloud, thereby facilitating its utilisation by theoretical scientists and materials companies \cite{Ichikawa2024NatRev}.  
This has led to a proliferation of research papers and significant advancements in this field \cite{Ishiyama2022SciRep, Weaving2023PRR, Xie2024PRA, Selisko2024arXiv, Robledo2024arXiv, Seki2024arXiv}.

On the other hand, spin defects in wide-gap semiconductors are attracting attention as promising quantum sensing materials because of their high sensitivity to magnetic fields, capability of room-temperature operation, and tolerance to extreme environments, such as radiation and high temperatures and pressures. 
In particular, the negatively charged nitrogen vacancy, the so-called  $\mathrm{NV}^-$ center in diamond \cite{Buckley2010Science, Philipp2010Science}, which exhibits the highest magnetic field sensitivity, has already been reported to have various applications, such as magnetoencephalography and temperature measurement of local regions within cells. 
Silicon carbide (4H-SiC), which is increasingly being applied as a power device \cite{Kimoto, Matsunami1997, Baliga1989IEEE, Kobayashi2020APE}, is also known to be able to create qubits compatible with room-temperature operation \cite{Weber2010PNAS, Weber2011JAP, Koehl2011Nature, Falk2013NatComm, Gordon2013MRS, Widmann2015NatMat} and is attracting attention because of its ability to manufacture and process high-purity crystals. 
Similarly, ultrahigh-purity crystal growth \cite{Carlos2001JCG, Matthias2010PSSC, Okumura2011APE} and doping \cite{Yoshitaka2004APL} are possible with wurtzite-type aluminum nitride (w-AlN), which has a wide band gap i.e., 6.12 eV, \cite{Li2003APL} and is attracting attention as a host material for spin qubits. 
First-principles calculations have theoretically predicted that complex defects such as $(\mathrm{Zr_{Al}V_{N}})^{0}$, $(\mathrm{Hf_{Al}V_{N}})^{0}$, and $(\mathrm{Ti_{Al}V_{N}})^{0}$ are promising candidates for spin qubits \cite{Varley2016PRB, Seo2017PRM}.

In this study, we employed the probabilistic imaginary-time evolution (PITE) \cite{Kosugi2022PRR}, an algorithm that has been demonstrated to provide quantum acceleration for ground state preparation \cite{Nishi2024PRR}, to simulate the electronic structure of spin defects on Quantinuum trapped-ion quantum computer \cite{Pino2021Nature, Ryan2022arXiv, Self2024NPhys}. 
In order to account for spin defects in the crystals, including $\mathrm{NV}^{-}$ center in diamond and complex defects in w-AlN \cite{Varley2016PRB, Seo2017PRM}, we constructed a low-energy effective Hamiltonian \cite{Aryasetiawan2004PRB, Imada2010JPSJ}.
It is established that the singlet states of $\mathrm{NV}^{-}$ center in diamond are multi-configuration states, rendering them challenging to describe using the density functional theory (DFT) \cite{Hohenberg1964, Kohn1965} framework, which is formulated based on a single Slater determinant.
The PITE method was employed for the low-energy effective Hamiltonian, thereby enabling an accurate description of the singlet states of the NV$^{-}$ center in diamond. This methodology will facilitate the discovery of a new spin defect for quantum sensing materials.

Quantinuum H1-1 computer has high-fidelity two-qubit gate of $\sim 2.0\times 10^{-3}$ infidelity with the all-to-all connectivity, performs mid-circuit measurement, and presents quantum volume of $2^{20}$.
However, the remaining quantum errors may worsen the computed results, and we address this issue by implementing quantum error detection (QED) \cite{Steane1996PRA, Gottesman1998PRA, Knill2004arXiv, Knill2004arXiv2, Self2024NPhys}.
In particular, we employed the Iceberg code \cite{Self2024NPhys}, which requires only four ancilla qubits for encoding and syndrome measurements. 
While the QED code is unable to reproduce the noiseless results and must therefore discard the erroneous results, it does, however, permit the implementation of the QED code on current computers. 
In QED, ancilla qubits are employed for each cycle of syndrome measurement, similarly in PITE, ancilla qubits are utilised for each cycle of the small imaginary time evolution.
The feature of the Quantinuum H1-1 computer, which performs mid-circuit measurement and reuse, is desirable for executing QED and PITE.

\section{method}
\label{sec:method}
\subsection{An low-energy effective model of the spin-defect qubits}
\label{sec:method_classical}
\subsubsection{DFT calculation}
DFT calculations were performed using a plane-wave basis set and optimized norm-conserving Vanderbilt pseudopotentials \cite{Troullier1991PRB, Hamann2013PRB, Schlipf2015CPC} as implemented in  the Quantum ESPRESSO code \cite{Giannozzi2009JPCM, Giannozzi2017JPCM, Giannozzi2020JCP}.
First, the structure of the complex defects in the supercell was optimized using spin-polarized DFT calculations with Perdew-Burke-Ernzerhof (PBE) semilocal functional \cite{Perdew1996PRL} until all forces were smaller than $10^{-3}$ a.u..  
As a result of confirming convergence in Appendix \ref{appendix:dft_calculation}, we adopted an energy cutoff of 50 Ry and a supercell of 216 atoms for the NV$^{-}$ center in diamond, and an energy cutoff of 75 Ry and a supercell of 240 atoms for the complex defects in w-AlN.
We confirmed the convergence of the energy cutoff within 113 meV and 1.9 meV for the NV$^{-}$ center in diamond and complex defects in w-AlN by calculating the same system with an energy cutoff of 100 Ry, respectively.
Brillouin zone, we employed $\Gamma$ point in both systems.
We employed $\Gamma$ point sampling for the Brillouin zone in both systems.
To obtain the wave function for constructing a low-energy effective model, DFT calculations with the screened hybrid functional of Heyd-Scuseria-Ernzerhof (HSE) \cite{Heyd2003JCP, Heyd2006JCP} were performed.

\subsubsection{A low-energy effective model}
In this study, in order to accurately describe phenomena in the low-energy region, we construct the extended Hubbard model given by 
\begin{gather}
    \mathcal{H}
    =
    \sum_{\sigma}\sum_{ij} t_{ij} a_{i\sigma}^{\dagger} a_{j\sigma}
    +
    \frac{1}{2} \sum_{\sigma \rho} \sum_{ij}
    \Big [
    U_{ij} a_{i\sigma}^{\dagger} a_{j\rho}^{\dagger} a_{j\rho} a_{i\sigma}
    \notag \\
    +
    J_{ij} (
        a_{i\sigma}^{\dagger} a_{j\rho}^{\dagger} a_{i\rho} a_{j\sigma}
        +
        a_{i\sigma}^{\dagger} a_{i\rho}^{\dagger} a_{j\rho} a_{j\sigma}
    )
    \Big],
\label{eq:extended_hubbard_hamiltonian}
\end{gather}
where $a_{i\sigma}^{\dagger}$ and $a_{i\sigma}$ denote the creation and annihilation operators, respectively, of an electron with spin $\sigma$ in the $i$th Wannier orbital, respectively.
A transfer integral $t_{ij}$ is defined as
\begin{gather}
    t_{ij}
    =
    \langle \phi_{i} | \mathcal{H}_0 | \phi_j \rangle
\end{gather}
where $|\phi_i\rangle$ denotes the spatial part of the wave function created by $a_{i}^{\dagger}$.
$\mathcal{H}_0$ represents the one-body Hamiltonian, here we take it as Kohn-Sham (KS) Hamiltonian $\mathcal{H}_{\mathrm{KS}}$ computed in previous subsection.
The screened-Coulomb $U_{ij}$ and exchange $J_{ij}$ integrals are given by
\begin{gather}
    U_{ij}(\omega)
    =
    \langle \phi_i \phi_j | W(\omega) |\phi_i \phi_j\rangle
\end{gather}
and
\begin{gather}
    J_{ij}(\omega)
    =
    \langle \phi_i \phi_j | W(\omega) |\phi_j \phi_i\rangle,
\end{gather}
where $W(\omega)$ denotes the frequency-dependent screened Coulomb interaction based on constrained random phase approximation (cRPA).
The static limit of the screened Coulomb integrals is used to describe the interaction in the Hamiltonian:
\begin{gather}
    U_{ij}
    =
    \lim_{\omega \to 0}
    U_{ij}(\omega)
\end{gather}
and
\begin{gather}
    J_{ij}
    =
    \lim_{\omega \to 0}
    J_{ij}(\omega).
\end{gather}
Note that a methodology to remove double counting of two-electron integrals from the one-electron integral was proposed \cite{Ido2024CPC}; however, we ignored double counting because the effect on the full-configuration interaction (FCI) energy is negligible in this study.

In this study, the direct Coulomb and exchange integrals were calculated using the software RESPACK \cite{Nakamura2021CPC}.
We used the cutoff energy for the polarization function $E_{\mathrm{cut}}^{\epsilon}$ of 3.0Ry and 4.5Ry for $\mathrm{NV}^{-}$ center in diamond and complex defects in w-AlN, respectively. We used KS energy levels $N_{\mathrm{band}}$ of 750 and 1300 for $\mathrm{NV}^{-}$ center in diamond and complex defects in w-AlN, respectively. 
In Appendix \ref{appendix:dft_calculation}, it is confirmed that the energy of the FCI calculation converges by taking the value $E_{\mathrm{cut}}^{\epsilon}$ of 6\% of the energy cutoff of the DFT calculation and by including conduction bands greater than 10 eV from the valence band maximum (VBM).

We define a linear transformation between Wannier orbitals and KS orbitals such that
$
    a_{i}
    =
    \sum_{j} U_{i, j} c_{j},
$
where $c_{j}$ denotes the annihilation operator of the KS-basis.
$j$ for $c_j$ is the composite notation for the wave vector and band index of the KS basis.
Using the linear transformation, the Hamiltonian based on the KS orbitals is given by
\begin{gather}
    \mathcal{H}
    =
    \sum_{\sigma} \sum_{i, j}
    \varepsilon^{\mathrm{KS}}_{i}
    c_{i\sigma}^{\dagger} c_{i\sigma}
    +
    \frac{1}{2}
    \sum_{\sigma \rho} \sum_{i, j, k, l} 
    \tilde{v}_{ijkl} 
    c_{i\sigma}^{\dagger} c_{j\rho}^{\dagger}
    c_{l\rho} c_{k\sigma}.
\label{eq:hamiltonian_KS_orbital}
\end{gather}
Each interaction term in Eq. (\ref{eq:hamiltonian_KS_orbital}) involves four distinct KS orbitals, leading to $O(n^4)$ scaling for $n$ Wannier orbitals. 
This is in contrast to the original Hamiltonian based on the Wannier basis in  Eq. (\ref{eq:extended_hubbard_hamiltonian}) with $O(n^2)$ scaling.
% The extended Hubbard model based on the Wannier-basis has an $O(n^2)$ scaling with Wannier orbitals $n$, which is advantageous for quantum circuit implementation from the perspective of scaling \cite{Yoshida2024arXiv}.
As described below, we used the $\mathbb{Z}_2$ symmetry to reduce the number of qubits necessary. However, the $\mathbb{Z}_2$ symmetry cannot be used for the Wannier-basis representation; therefore, we converted it to the KS-basis representation in this study.

\subsubsection{Spin operator representation}

To implement the ITE of fermionic operators on a quantum computer, the transformation of fermionic operators to Pauli operators is necessary \cite{Jordan1928, Bravyi2002AnnPhys, Jacob2012JCP}:
\begin{gather}
    \mathcal{H}
    =
    \sum_{\ell = 1}^{N_{\mathrm{term}}} h_{\ell} P_{\ell} ,
\label{eq:hamiltonian_based_on_pauli}
\end{gather}
where $h_{\ell}$ denotes the real coefficients, and $P_{\ell}$ is the Pauli operator acting on some of $n$ qubits.
Here, we utilize parity transformation \cite{Jacob2012JCP} for the low-energy effective Hamiltonian based on the KS orbitals.
By indicating the parity of the total number of up and down spins, the parity transformation can reduce the number of qubits necessary by two.

Further qubit reduction is achieved by using $\mathbb{Z}_2$ symmetry\cite{Bravyi2017arXiv}. 
This function identifies a Pauli operators $P_{Z_2}$ such that 
$
    [P_{Z_2}, \mathcal{H}] =0
$.
If we find such a Pauli operator, the eigenstates of the Pauli operator can be classified as either +1 or $-1$ eigenspace.
In other words, it becomes possible to reduce the degree of freedom by one qubit.
%In this study, we could successfully discover a $\mathbb{Z}_2$ symmetry by ignoring terms with small coefficients in the Hamiltonian, reducing the number of qubits by 1.

\subsection{Probabilistic imaginary-time evolution with Iceberg code}
\subsubsection{Probabilistic imaginary-time evolution}
% where the subscript $L$ indicates encoded states
In this study, we adopted the PITE method \cite{Kosugi2022PRR} to obtain the lowest energy state among eigenvectors included in an initial state.
After determining the ground state, the ground-state energy can be efficiently estimated using QPE \cite{Kitaev1995arXiv, Abrams1997PRL, Omalley2016PRX, Paesani2017PRL, Lin2022PRXQ, Yamamoto2024PRR}.
For the Hamiltonian $\mathcal{H}$ for an $n$-qubit system, we define a nonunitary Hermitian operator $\mathcal{M} := m_0 e^{-\mathcal{H}\Delta\tau}$, where $\Delta\tau$ denotes a small imaginary time-step size, and an adjustable real parameter $m_0$ satisfying $0 < m_0 < 1$, $m_0 \neq 1/\sqrt{2}$ is introduced. 
Introducing an ancilla qubit, the nonunitary operator is embeded as a submatrix into an extended unitary matrix as  
\begin{gather}
    \mathcal{U}_{\mathcal{M}}
    =
    \begin{pmatrix}
        \mathcal{M} & \sqrt{1 - \mathcal{M}^2} \\
        \sqrt{1 - \mathcal{M}^2} & -\mathcal{M}
    \end{pmatrix} ,
\end{gather}
where the unitary matrix $\mathcal{U}_{\mathcal{M}}$ is divided into submatrices depending on the state of the ancilla qubit (the basis of the top left is $|0\rangle\langle 0|$).
Operating the extended unitary operator on an input state prepared as $|\psi\rangle |0\rangle$,
we have an entanglement state of the nonunitary operator acted (success state) and other state (failure state):
\begin{gather}
    \mathcal{U}_{\mathcal{M}}
    |\psi\rangle|0\rangle
    =
    \mathcal{M} |\psi\rangle|0\rangle
    +
    \sqrt{1 - \mathcal{M}^2}|\psi\rangle |1\rangle .
\end{gather}
Measuring the ancilla qubit as $|0\rangle$ state  with probability $\mathbb{P}_0 = \langle \psi |\mathcal{M}^2 | \psi\rangle$
collapses the entanglement state into the success state, 
$
    (1/\sqrt{\mathbb{P}_0}) \mathcal{M}
    |\psi\rangle
$
.

\begin{figure}[h]
    \centering
    \includegraphics[width=0.45 \textwidth]{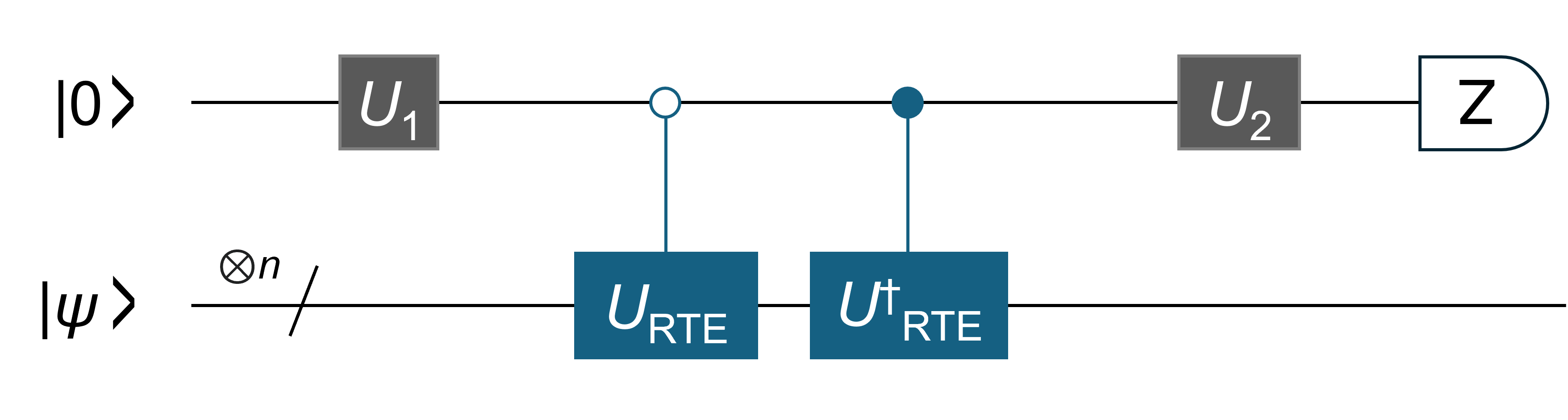} 
\caption{
Circuit diagram of the logical approximated PITE operator. 
}
\label{fig:qc_pite}
\end{figure}

Since it is difficult to decompose $\mathcal{U}_{\mathcal{M}}$ for a general Hamiltonian into single and two-qubit gates, a previous study \cite{Kosugi2022PRR} proposed an approximate implementation method for the nonunitary operator $\mathcal{M}$ within the first order of $\Delta\tau$, by executing Taylor expansion. 
The approximated ITE operator is given by
\begin{gather}
    \mathcal{M}
    =
    \sin(-\mathcal{H} \Delta\tau s_1 + \varphi) 
    +
    O(\Delta\tau^2),
\label{eq:approximated_pite}
\end{gather}
where $s_1 = m_0 / \sqrt{1 - m_0^2}$ and $\varphi = \tan^{-1} s_1$.
The quantum circuit for the approximated PITE  is shown in Figure \ref{fig:qc_pite}. 
The quantum circuit is comprising of two single-qubit unitary operators 
$
    U_1 = R_Z\left(\pi/2\right) R_Y(\pi) R_Z(-\pi) 
$
    and
$
    U_2 = R_Z\left(\pi/2\right) R_Y(\pi) R_Z(2\theta_0 - 3\pi/2)
$,
and forward- and backward-controlled real-time evolution (CRTE) operators,
$
    e^{ \Delta\tau s_1 \mathcal{H}} 
    \otimes |0\rangle\langle 0|
    +
    e^{- \Delta\tau s_1 \mathcal{H}} 
    \otimes |0\rangle\langle 0|
$
for time $\Delta\tau s_1$, where 
$   
    \theta_0 = \kappa \arccos [(m_0 + \sqrt{1 - m_0^2})/\sqrt{2}]
$ 
and
$\kappa = \mathrm{sgn} ( m_0 - 1 / \sqrt{2})$.

\subsubsection{Iceberg code}
\label{sec:Iceberg_code}
In this subsection, we briefly review the implementation of a $\llbracket n+2,n,2 \rrbracket$ QED code called the Iceberg code \cite{Self2024NPhys}.
The Iceberg code encodes $k$ even logical qubits $L := \{q_{0}, q_1, \ldots, q_{k-1}\}$ into $k+2$ physical qubits $T := D \cup A$, where $D = L$ and $A := \{q_X, q_Z\}$.
The data register $D$ stores the same quantum state as the logical qubit state of $L$, and the Iceberg code employs two redundant qubits $A$ for encoding.
The Iceberg code is classified as stabilizer code, and the code space is stabilized by $S_X = \otimes_{i\in T}X_i$ and $S_Z = \otimes_{i\in T}Z_i$.
A syndrome measurement using two additional ancilla qubits can detect a single-qubit error.
A $k$-bit string state $|x\rangle_L = |x_{k-1}\rangle \ldots |x_1\rangle |x_0\rangle$ is encoded as 
\begin{gather}
    |x\rangle_L
    =
    \frac{
        |0 \rangle_{q_Z}
        |f_x \rangle_{q_X} 
        |x\rangle_D
        + 
        |1 \rangle_{q_Z}
        |\neg f_x \rangle_{q_X} 
        |\neg x\rangle_D
    }{\sqrt{2}} ,
\end{gather}
where $\neg x$ is the logical not (negation) of $x$ in the binary representation, $f_x = 0$ for even ${\sum_{i=0}^{k-1}x_i}$, and $f_x =1$ for odd ${\sum_{i=0}^{k-1}x_i}$.

In the Iceberg code, the logical Pauli-$X$ gates are defined as
\begin{gather}
\begin{aligned}
    \overline{X}_i
    &=
    X_{q_X} X_i 
    ~~ \forall i \in L 
    \\
    \overline{X}_i \overline{X}_j
    &=
    X_i X_j 
    ~~ \forall i, j \in L
    \\
    \otimes_{j \in L \backslash i} \overline{X}_j
    &=
    X_{q_Z} X_i 
    ~~ \forall i L 
    \\
    \otimes_{j \in L} \overline{X}_j
    &=
    X_{q_X} X_{q_Z} ,
\label{eq:logical_pauli_x}
\end{aligned}
\end{gather}
and the logical Pauli-$Z$ gates are
\begin{gather}
\begin{aligned}
    \overline{Z}_i
    &=
    Z_{q_Z} Z_i 
    ~~\forall i \in L
    \\
    \overline{Z}_i \overline{Z}_j
    &=
    Z_i Z_j 
    ~~\forall i, j \in L
    \\
    \otimes_{j \in L \backslash i} \overline{Z}_j
    &=
    Z_{q_X} Z_i 
    ~~\forall i \in L 
    \\
    \otimes_{j \in L} \bar{Z}_j
    &=
    Z_{q_X} Z_{q_Z} .
\label{eq:logical_pauli_z}
\end{aligned}
\end{gather}
Also, the logical Pauli-$Y$ gates are given by
\begin{gather}
\begin{aligned}
    \overline{Y}_i \overline{Y}_j
    &=
    Y_i Y_j 
    ~~\forall i, j \in L 
    \\
    \overline{X}_i \otimes_{j \in L \backslash i} \overline{Z}_j
    &=
    -Y_{q_X} Y_i 
    ~~\forall i, j \in L 
    \\
    \overline{Z}_i \otimes_{j \in L \backslash i} \overline{X}_j
    &=
    -Y_{q_Z} Y_i 
    ~~ \forall i \in L
    \\
    \otimes_{j \in L} \overline{Y}_j
    &=
    (-1)^{1+k / 2} Y_{q_X} Y_{q_Z} .
\label{eq:logical_pauli_y}
\end{aligned}
\end{gather}
Here, the overline on an operator is used to clearly indicate that the operator is a logical gate.

A previous study \cite{Self2024NPhys} introduced the implementation method of logical rotation gates in a transversal manner using M\o lmer-S\o rensen (MS) gate \cite{Molmer1999PRL}. The definition of the MS gate is
\begin{gather}
    \mathrm{MS}_{ij}(\theta)
    =
    \exp\left(
        -i \frac{\theta}{2}
        Z_i \otimes Z_j
    \right) .
\end{gather}
Logical rotation gates 
$
    \overline{R}_{P}(\theta)
    :=
    e^{
        -i \frac{\theta}{2}
        \overline{P}
    }
$
with respect to Pauli operators $\overline{P}$ given by Eqs. (\ref{eq:logical_pauli_x}), (\ref{eq:logical_pauli_z}), and (\ref{eq:logical_pauli_y}) can be implemented using an MS gate and a single-qubit gate for a basis transformation. 
For example, $\overline{R}_{XX}(\theta)$ is implemented as 
$
    \overline{R}_{XX}(\theta)
    =
    (H \otimes H)
    \mathrm{MS}(\theta)
    (H \otimes H)
$.
Using the logical Pauli gates and logical rotation gates defined above, universal logical computation can be performed. Note that the encoding and syndrome measurements are performed in a fault-tolerant manner in the Iceberg code; however, the logical rotation gate is compiled into the physical gates set in a non-fault-tolerant manner.

\subsubsection{Overall circuit}
\begin{figure*}
    \centering
    \includegraphics[width=0.95 \textwidth]{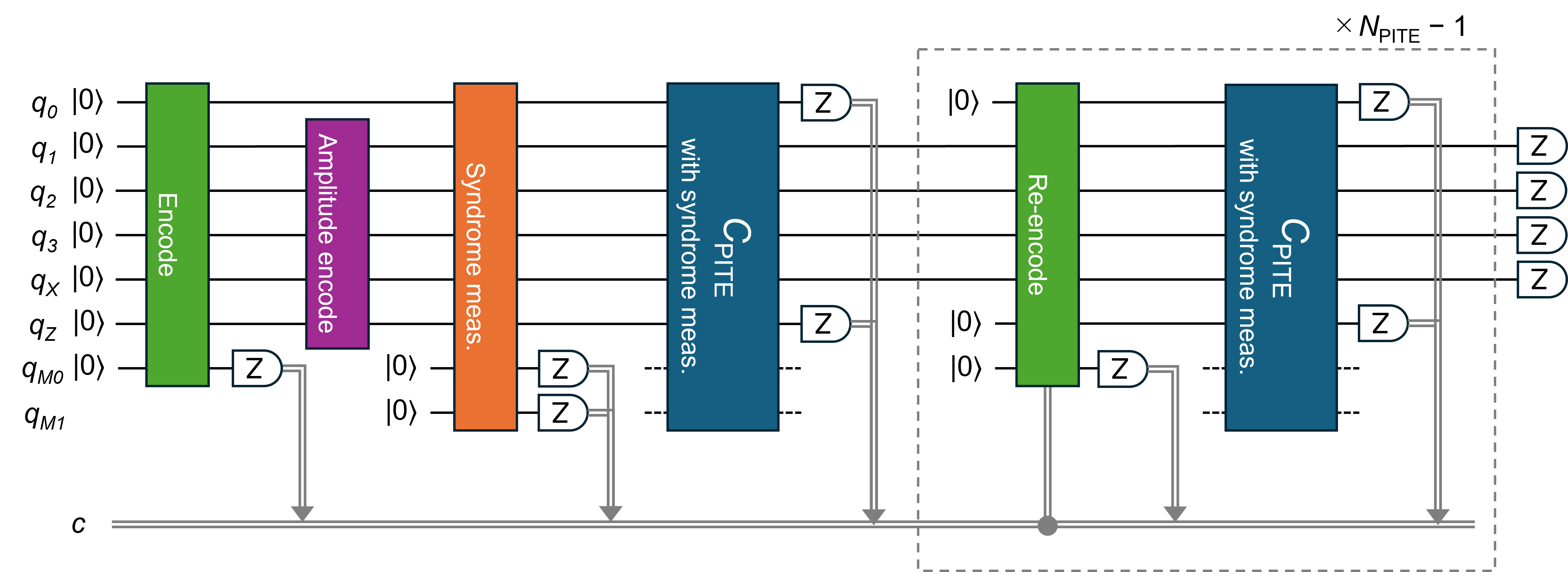} 
\caption{
The entire encoded quantum circuit for the PITE comprises eight quantum registers $\{q_i\}$ and classical registers $c$.
The measurement results with respect to $Z$ basis are stored in the classical registers $c$ (gray lines with arrows). 
The encoded circuit generates $|0\rangle_{L}$ in code space. 
The $\{q_i\}_{i=0,1,2,3}$ qubits are assigned to the data register, while the remaining qubits are designated as anciila qubits for the iceberg code. 
The $q_0$ qubit is assigned to an ancilla qubit within the PITE circuit.
The amplitude-encoding circuit is operated to prepare an initial state for the PITE, which contains approximately thirty two-qubit gates. 
Subsequently, the syndrome measurement is applied. 
After preparing the encoded initial state, the first step of the PITE is performed, including several syndrome measurements that act at almost equal intervals.
Then, $q_0$ and $q_Z$ registers are observed to collapse the entangled state to the success state. 
After measuring $q_0$ and $q_Z$ registers, the re-encode circuit is applied to recover the encoded state, which contains conditional operations based on classical registers. 
Re-encoding circuits, PITE circuits with syndrome measurement, $Z$-basis measurement are repeatedly performed until sufficient imaginary-time has elapsed.
}
\label{fig:circuit_pite}
\end{figure*}

\begin{figure*}
    \centering
    \includegraphics[width=0.95 \textwidth]{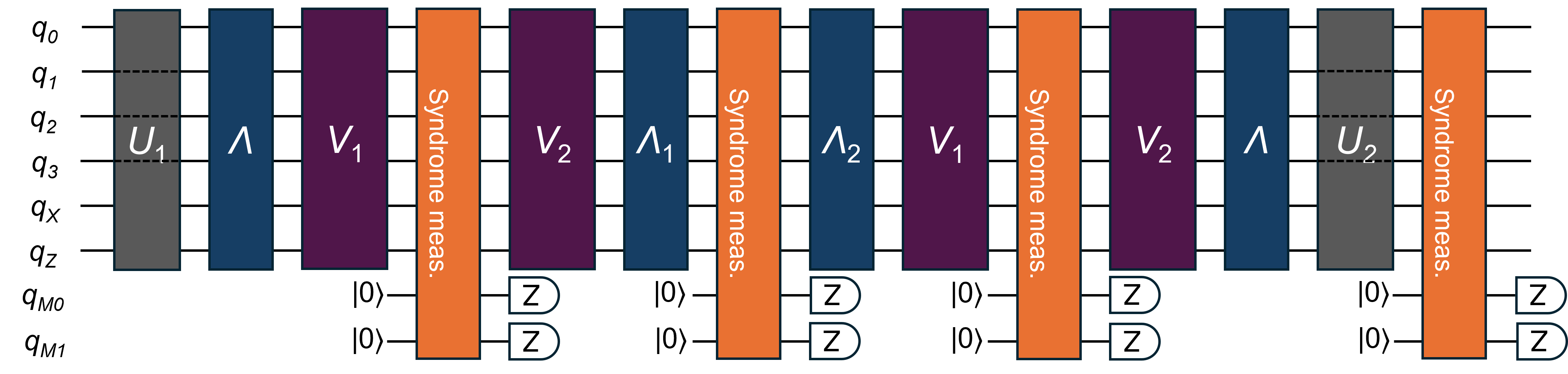} 
\caption{The detailed circuit diagram of PITE including syndrome measurements in Figure \ref{fig:circuit_pite}.
$U_1$ and $U_2$ denote the logical single-qubit gates on the ancilla qubit of PITE. Other blocks, excluding the syndrome measurement, are logical RTE operators, and the generators are indicated in these blocks.
The logical operation excluding the syndrome measurements, $U_1$ and $U_2$ in this figure is expressed as 
$
    e^{-i\frac{s\Delta \tau}{4}\Lambda}
    e^{-i\frac{s\Delta \tau}{2}V_2}
    e^{-i\frac{s\Delta \tau}{2}V_1}
    e^{-i\frac{s\Delta \tau}{2}\Lambda_2}
    e^{-i\frac{s\Delta \tau}{2}\Lambda_1}
    e^{-i\frac{s\Delta \tau}{2}V_2}
    e^{-i\frac{s\Delta \tau}{2}V_1}
    e^{-i\frac{s\Delta \tau}{4}\Lambda}
$.
}
\label{fig:circuit_ptie_single}
\end{figure*}

\begin{figure}
    \centering
    \includegraphics[width=0.45 \textwidth]{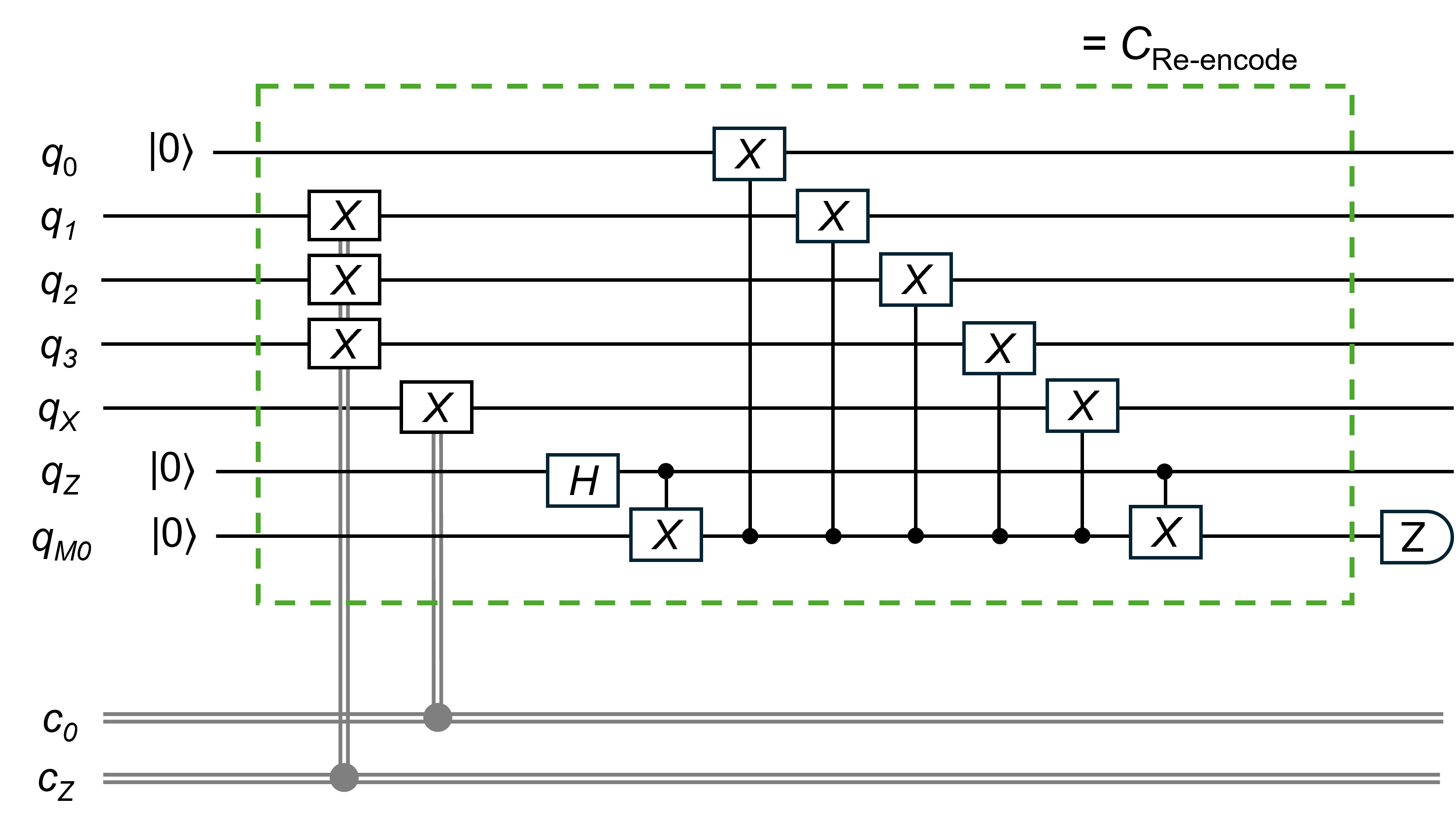} 
\caption{
Detailed circuit diagram of the re-encoding block $C_{\mathrm{Re-encode}}$ of the quantum circuit shown in Figure \ref{fig:circuit_pite}. Classical registers $c_0$ and $c_{Z}$ store the measurement results of quantum registers $q_0$ and $q_Z$, respectively.
}
\label{fig:circuit_pite_reencode}
\end{figure}

Although a recent study analyzed that the ITE method on a quantum computer was robust against quantum noise \cite{Khindanov2024arXiv}, the PITE method usually requires a large number of gate operations. 
Therefore, quantum devices that do not incorporate QED and quantum error correction (QEC) are undesirable for the PITE method. 
As a first step, we consider the implementation of the Iceberg code \cite{Self2024NPhys}. 
The entire encoded quantum circuit for the PITE is depicted in Fig. \ref{fig:circuit_pite}.
The quantum circuit used in this study consists of a total of eight qubits: six qubits for encoding a four-qubit system and two qubits ($q_{M0}$ and $q_{M1}$) for syndrome measurement.
The $\{q_i\}_{i=0,1,2,3}$ qubits are assigned to the data register.
Specifically, the $q_0$ qubit is assigned to the ancilla qubit within the PITE circuit.
First, an encoding circuit is applied to generate a logical $|0\rangle_{L}$ state \cite{Self2024NPhys}. 
A single ancilla qubit is used as a flagged qubit to detect errors that occur during encoding \cite{Chao2018PRL}.
In total, this involves seven two-qubit gate operations.
Amplitude encoding is used to generate an encoded initial state $|\psi\rangle_{L}$ for PITE, where approximately 30 physical two-qubit gates are required for logical operations for amplitude encoding. 
Therefore, we performed syndrome measurements after amplitude encoding.
Twelve two-qubit gates are required for the syndrome measurement.
The PITE quantum circuit is then executed, as shown in Fig. \ref{fig:circuit_ptie_single}. 
Because this quantum circuit involves many gate operations, a syndrome measurement is periodically performed within the block.
Subsequently, by measuring the $q_0$ and $q_Z$ quantum registers, the success state is obtained.
The logical state is broken by measuring $q_0$ and $q_Z$ registers. Thus, we reset the $q_0$ and $q_Z$ registers and perform a re-encoding block to regenerate the encoded state. 
A detailed circuit diagram of the re-encoding block is presented in Fig. \ref{fig:circuit_pite_reencode}, where the flagged qubit is also used. 
The next PITE block is executed in the re-encoded state.
This procedure is repeatedly applied until sufficient imaging time has elapsed. 
Finally, we performed destructive measurements of the remaining quantum registers.

\subsubsection{PITE in encoded state}
The logical amplitude encoding is implemented by decomposing it into native gates of the Iceberg code such as $\{R_{X}, R_{Z}, R_{XX}, R_{YY}, R_{ZZ}\}$.
However, to optimize the number of gate operations, we consider a detailed implementation of the CRTE gate, which is the main building block of our algorithm.
The main feature of the Iceberg code is that it can efficiently implement global rotations that act on the entire logical space.
The forward- and backward-CRTE gates in the approximated PITE circuit are rewritten as
\begin{gather} 
    e^{-is\Delta\tau \mathcal{H}}
    \otimes
    |0\rangle \langle 0|
    +
    e^{is\Delta\tau \mathcal{H}}
    \otimes
    |1\rangle \langle 1|
    =
    e^{
        -i s\Delta\tau
        \mathcal{H} \otimes Z
    }.
\end{gather}
As shown in Eq. (\ref{eq:hamiltonian_based_on_pauli}), because the Hamiltonian is expressed as a linear combination of Pauli operators, the CRTE operator can be decomposed into a sequence of rotation gates using Trotter-Suzuki decomposition \cite{Suzuki1991JMP}.
Let us now consider an efficient implementation. First, we decompose the generator of the CRTE operators as
\begin{gather}
    \mathcal{H} \otimes Z
    =
    \Lambda + V, 
\label{eq:generator_of_crte}
\end{gather}
where $\Lambda$ denotes the diagonal part and $V$ is the non-diagonal part of the generator.
Importantly, the computational model in this study comprises of the larger values of diagonal coefficients for $\Lambda$ and smaller values of non-diagonal coefficients for $V$. This is often the case with quantum chemistry, because you usually start from a good mean-field Hamiltonian such as DFT or Hartree-Fock to include the electron correlation. We would like to stress here that by introducing such a  decomposition, we can implement $\Lambda+V$ efficiently, as follows.
We used the second-order Trotter-Suzuki decomposition as:
\begin{gather}
    e^{
        -i t (\Lambda + V)
    }
    =
    \left[
    e^{
        -i \frac{t}{2r} \Lambda
    }
    e^{
        -i \frac{t}{r} V
    }
    e^{
        -i \frac{t}{2r} \Lambda
    }
    \right]^{r}
    +
    O\left[\left(
        \frac{t}{r}
    \right)^{3}\right] .
\end{gather}
Note that we can implement the RTE of $\Lambda$ exactly.
The first-order Trotter-Suzuki decomposition is sufficient to implement the RTE for $V$ composed of the small coefficients of the Pauli operators.
% We stress that the additional cost for implementing only $e^{it\Lambda /(2r)}$ operator leads to reduction of the error arising from non-commutator from first-order to second-order of $t$ in the second-order Trotter-Suzuki decomposition.
We remark that the second-order Trotter decomposition can be implemented only for the additional cost of $e^{it\Lambda /(2r)}$ to the first-order Trotter decomposition.
Figure \ref{fig:circuit_ptie_single} shows a detailed schematic of the PITE circuit when the dividing number of the CRTE gate is $r=2$.
The quantum circuit was implemented by decomposing it as $\Lambda = \Lambda_1 + \Lambda_2$ and $V = V_1 + V_2$, so that the syndrome measurement is inserted after every interval of approximately 40 quantum gate operations.
The specific expression of Eq. (\ref{eq:generator_of_crte}) used in this study is given in the appendix \ref{sec:model_paremeters}.

% \subsection{Intermediate measurements of the logical qubit in PITE}
The input state for the PITE is prepared for the quantum register $L_S := \{q_1, q_2, q_3\}$.
The encoded initial state $|\psi\rangle_{L_S} |0\rangle_{q_0}$ is changed by applying PITE operator as,
\begin{gather}
    \sum_{k=0}^{N-1}
    \left(
        c^{(\mathrm{succ})}_k |k\rangle_{L_S} |0\rangle_{q_0}
        +
        c^{(\mathrm{fail})}_k |k\rangle_{L_S} |1\rangle_{q_0}
    \right) ,
\end{gather}
where $N=2^n$, and ${c_k^{(\mathrm{succ})}}$ and ${c_k^{(\mathrm{fail})}}$ represent the expansion coefficients of the success and failure states with respect to the computational basis, respectively.
In the Iceberg code, the state acted on by the ITE operator is represented by the physical qubits as 
\begin{gather}
    \sum_{k=0}^{N-1}
    c^{(\mathrm{succ})}_k 
    \frac{
        |0\rangle_{q_Z} |f_k \rangle_{q_X}  |k\rangle_{D_S} |0\rangle_{q_0}
        +
        |1\rangle_{q_Z} |\neg {f_k} \rangle_{q_X} |\neg k \rangle_{D_S} |1 \rangle_{q_0}
    }{\sqrt{2}}
    \notag \\
    +
    c^{(\mathrm{fail})}_k 
    \frac{
        |0\rangle_{q_Z} |\neg {f_k} \rangle_{q_X} |k\rangle_{D_S} |1\rangle_{q_0}
        +
        |1\rangle_{q_Z} |f_k \rangle_{q_X} |\neg k\rangle_{D_S} |0 \rangle_{q_0}
    }{\sqrt{2}} .
\end{gather}
Measuring $q_0$ and $q_Z$ collapses the entangled state to the success state:
\begin{gather}
    \frac{1}{\sqrt{p_{\mathrm{succ}}}}
    \sum_{k=0}^{N-1}
    c^{(\mathrm{succ})}_k 
    \frac{
        |f_k \rangle_{q_X} |k\rangle_{D_S}
    }{\sqrt{2}} ,
\end{gather}
when observing the physical qubits as $|0\rangle_{q_Z}|0\rangle_{q_0}$ state and 
\begin{gather}
    \frac{1}{\sqrt{p_{\mathrm{succ}}}}
    \sum_{k=0}^{N-1}
    c^{(\mathrm{succ})}_k 
    \frac{
          |\neg f_k \rangle_{q_X} |\neg k \rangle_{D_S}
    }{\sqrt{2}} ,
\end{gather}
for the physical qubit state being $|1\rangle_{q_Z}|1\rangle_{q_0}$ state.
Here, $p_{\mathrm{succ}}$ is the success probability, where the encoded ancilla qubit of the PITE circuit is observed as $|0\rangle_{q_0}$ state.
In contrast, observing the unencoded qubits as $|0\rangle_{q_Z}|1\rangle_{q_0}$ and $|1\rangle_{q_Z}|0\rangle_{q_0}$ corresponds to the failure state.
If the ancilla qubit of $q_Z$ is measured as $|0\rangle_{q_Z}$ state, we operate $X$ gates on the system qubits. In addition, if the ancilla qubit of $q_0$ is measured as $|0\rangle_{q_0}$ state, we act $X$ gate on the ancilla qubit of $q_{X}$. 
By applying such conditional-$X$ gates with respect to classical registers, we have the unencoded states as 
\begin{gather}
    \sum_{k=0}^{N-1} \widetilde{c}^{(\mathrm{succ})}_k 
    |0\rangle_{q_Z} |f_k\rangle_{q_X}  |k\rangle_{D_S} |0\rangle_{q_0}
\end{gather}
for success state and
\begin{gather}
    \sum_{k=0}^{N-1} \widetilde{c}^{(\mathrm{fail})}_k 
    |0\rangle_{q_Z} |f_k\rangle_{q_X}  |k\rangle_{D_S} |0\rangle_{q_0}
\end{gather}
for failure state, where we reset the measured qubits of $q_0$ and $q_Z$ and 
$
    \widetilde{c}_k^{(\mathrm{succ})} 
    = 
    c_k^{(\mathrm{succ})} / \sqrt{p_{\mathrm{succ}}}
$.
The gate operations are shown in Fig. \ref{fig:circuit_pite_reencode} leads to the encoded state:
\begin{gather}
    \sum_{k=0}^{N-1} \widetilde{c}^{(\mathrm{succ})}_k 
    \frac{
        |0\rangle_{q_Z} |f_k\rangle_{q_X} |k\rangle_{D_S} |0\rangle_{q_0}
        +
        |1\rangle_{q_Z} |\neg f_k\rangle_{q_X} |\neg k\rangle_{D_S} |1\rangle_{q_0}
    }
    {\sqrt{2}}
    \notag \\
    =
    \sum_{k=0}^{N-1} \widetilde{c}^{(\mathrm{succ})}_k 
    |k\rangle_{L_S} |0\rangle_{q_0}  .
\end{gather}
With the same manner as the encoded success state, the encoded failure state 
$
    \sum_{k=0}^{N-1} \widetilde{c}^{(\mathrm{fail})}_k 
    |k\rangle_{L_S} |0\rangle_{q_0} 
$
is obtained using the re-encoding circuit shown in Fig. \ref{fig:circuit_pite_reencode}.

\section{Results}
\label{sec:results}
% \subsection{Excited energy of the spin-qubit}
\subsection{Classical computation}
\label{sec:results_classical}
\subsubsection{NV center in diamond}
\begin{figure}
    \centering
    \includegraphics[width=0.45 \textwidth]{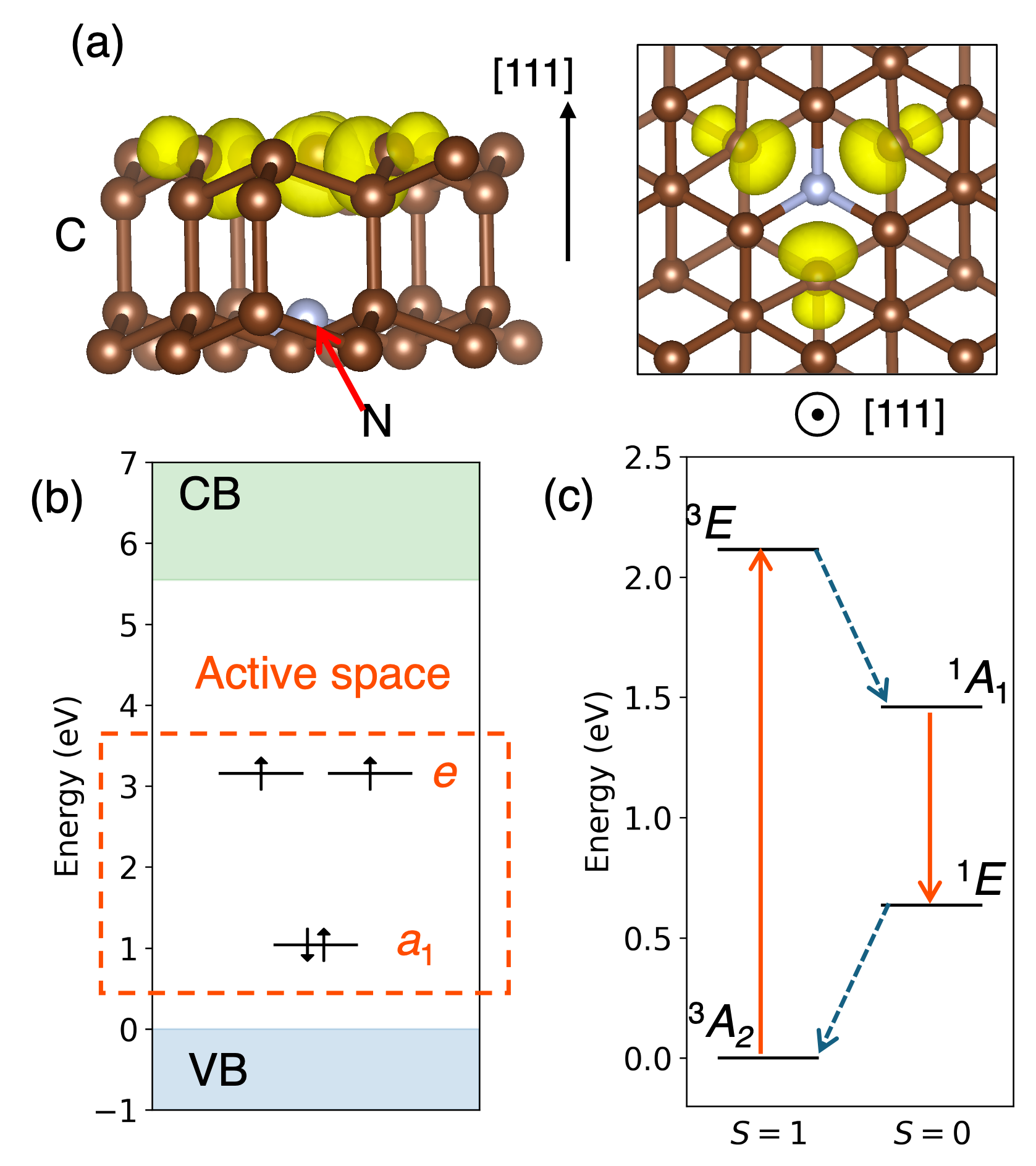} 
\caption{
The electronic structure of $\mathrm{NV}^{-}$ in diamond was computed on a classical computer.
(a) Spin densities obtained from spin-unrestricted DFT calculations.
Isosurfaces are displayed at 10\% of the maximum value.
(b) Single-particle defect levels obtained by spin-restricted DFT calculations. The orange dashed box denotes the energy region to be considered for constructing a low-energy effective Hamiltonian. 
(c) Excitation energies calculated using exact diagonalization (FCI calculations). 
}
\label{fig:dft_nv_diamond}
\end{figure}

The accuracy of the treatment of spin defects is verified by performing exact diagonalization (FCI calculation) of the derived low-energy effective model on a classical computer before quantum calculations are performed.
First, calculations are performed on $\mathrm{NV}^{-}$-center in diamond, which are often applied to high-accuracy calculation methods for spin defects.
Figure \ref{fig:dft_nv_diamond}(a) shows the spin density of the defect level obtained from the DFT calculations without spin restrictions. 
Figure \ref{fig:dft_nv_diamond}(b) shows the KS defect levels obtained from spin-restricted DFT calculations. 
This confirms that defect levels existed between the conduction and valence bands. 
The active space was set to include only the energy levels of gap states.  
It can be seen that the spin density with $C_{3v}$ symmetry is formed by the dangling bonds around the carbon vacancy directly above the [111] direction of the nitrogen atom, which results in the degeneracy of $e_x$ and $e_y$ orbitals in the KS defect levels.

The results of diagonalizing the low-energy effective model, which incorporates the effects of bands outside the active space as screening effects, are shown in Fig. \ref{fig:dft_nv_diamond}(c). 
This correctly reproduces the order of excited states, such as ${}^3A_{2}<{}^1E<{}^1A_1<{}^3E$. 
The eigenvectors of the effective Hamiltonian are summarized in Table \ref{table:wave_function_nv}.
By checking each spin state, ${}^3A_{2}$ and ${}^3E$ states correspond to spin triplet states, and ${}^1E$ and ${}^1A_1$ state correspond to spin singlet states. As is well known, looking at the Table \ref{table:wave_function_nv}, spin triplet states are described by single-Slater determinants. In contrast, spin singlet states in nature require a multiplicity of Slater determinants. 
The eigenvalues were consistent with those of previous theoretical and experimental studies.
In a previous study \cite{Bockstedte2018npjQM} which created an effective model at the level of cRPA, the excitation energy was 0.47/1.36/2.05 eV for ${}^1E/{}^{1}A_{1}/{}^{3}E$ states, which is in good agreement with the results of this study, which are summarized in Table \ref{table:excitation_energy_nv_diamond}. 
% ${}^1E<{}^1A_1<{}^3E=0.47<1.36<2.05$ eV
Ma et al. applied density matrix embedding theory (DMET) \cite{Knizia2012PRL, Wouters2016JCTC, Hung2020JCTC} that incorporates an exchange-correlation effect beyond the RPA approximation to the spin defects, and reported that it shows good agreement with experimental values as 0.56/1.76/2.00 eV for ${}^1E/{}^{1}A_{1}/{}^{3}E$ states \cite{Ma2020npjCM, Ma2021JCTC, Huang2022PRXQ}.
%  ${}^1E<{}^1A_1<{}^3E=0.56<1.76<2.00$ eV 
A comparison of the computational methods proposed thus far for calculating the excited states in $\mathrm{NV^{-}}$ center diamonds with the results of this study is summarized in Appendix \ref{sec:comparison_other_method}.

\begin{table}[ht]
    \centering
    \renewcommand{\arraystretch}{1.5} % Adjust row height
    \setlength{\tabcolsep}{6pt} % Adjust column spacing
    \newcolumntype{Y}{>{\centering\arraybackslash}X}
    \caption{Computed vertical excitation energies (eV) of $\mathrm{NV}^{-}$-center in diamond.}
    \begin{tabularx}{\columnwidth}{|Y|YY|}
        \hline
        % \rowcolor{gray}
        Excitation & This work & Expt. \\
        \hline
        ${}^3A_2 \to {}^3E$   & 2.11 & 1.945 \cite{Davies1976} \\
        ${}^3A_2 \to {}^1A_1$ & 1.46 &  \\
        ${}^3A_2 \to {}^1E$   & 0.64 & \\
        ${}^1E   \to {}^1A_1$ & 0.82 & 1.190 \cite{Rogers2008NJP} \\
        ${}^1A_1 \to {}^3E$   & 0.66 & $\approx$ 0.4 \cite{Thiering2017PRB} \\
        \hline
    \end{tabularx}
    \label{table:excitation_energy_nv_diamond}
\end{table}

\begin{table*}
\centering
\renewcommand{\arraystretch}{1.5}
\caption{Computed many-body wave function of $\mathrm{NV}^{-}$-center in diamond.}
\begin{tabular}{|c|c|}
\hline
    State   & Many-body wave function  \\
    \hline 
    ${}^3{E}$ & 
    $
        |e_{y\uparrow}e_{y\downarrow}e_{x\uparrow}a_{1\uparrow}\rangle, ~~
        |e_{y\uparrow}e_{y\downarrow}e_{x\downarrow}a_{1\downarrow}\rangle, ~~
        0.71(
            |e_{y\uparrow}e_{y\downarrow}e_{x\downarrow}a_{1\uparrow}\rangle
            +
            |e_{y\uparrow}e_{y\downarrow}e_{x\uparrow}a_{1\downarrow}\rangle
        ),
    $
    \\
    & 
    $
        |e_{y\uparrow}e_{x\uparrow}e_{x\downarrow}a_{1\uparrow}\rangle, ~~
        |e_{y\downarrow}e_{x\uparrow}e_{x\downarrow}a_{1\downarrow}\rangle, ~~
        0.71(
            |e_{y\downarrow}e_{x\uparrow}e_{x\downarrow}a_{1\uparrow}\rangle
            +
            |e_{y\uparrow}e_{x\uparrow}e_{x\downarrow}a_{1\downarrow}\rangle
        )
    $
    \\ \hline
    ${}^1 A_{1}$ &
    $
        0.68(
            |e_{x\uparrow}e_{x\downarrow}a_{1\uparrow}a_{1\downarrow}\rangle
            +
            |e_{y\uparrow}e_{y\downarrow}a_{1\uparrow}a_{1\downarrow}\rangle
        )
        -
        0.26|e_{y\uparrow}e_{y\downarrow}e_{x\uparrow}e_{x\downarrow}\rangle
    $
    \\ \hline
    ${}^1{E}$ & 
    $
        0.51(
            |e_{y\uparrow}e_{x\downarrow}a_{1\uparrow}a_{1\downarrow}\rangle
            -
            |e_{y\downarrow}e_{x\uparrow}a_{1\uparrow}a_{1\downarrow}\rangle
        )
        +
        0.19(
            |e_{y\uparrow}e_{y\downarrow}e_{x\downarrow}a_{1\uparrow}\rangle
            -
            |e_{y\uparrow}e_{y\downarrow}e_{x\uparrow}a_{1\downarrow}\rangle
        )
        -
        0.45(
            |e_{x\uparrow}e_{x\downarrow}a_{1\uparrow}a_{1\downarrow}\rangle
            -
            |e_{y\uparrow}e_{y\downarrow}a_{1\uparrow}a_{1\downarrow}\rangle
        ),
    $
    \\
    & 
    $
        0.51(
            |e_{x\uparrow}e_{x\downarrow}a_{1\uparrow}a_{1\downarrow}\rangle
            -
            |e_{y\uparrow}e_{y\downarrow}a_{1\uparrow}a_{1\downarrow}\rangle
        )
        +
        0.19(
            |e_{y\downarrow}e_{x\uparrow}e_{x\downarrow}a_{1\uparrow}\rangle
            -
            |e_{y\uparrow}e_{x\uparrow}e_{x\downarrow}a_{1\downarrow}\rangle
        )
        -
        0.45(
            |e_{y\uparrow}e_{x\downarrow}a_{1\uparrow}a_{1\downarrow}\rangle
            -
            |e_{y\downarrow}e_{x\uparrow}a_{1\uparrow}a_{1\downarrow}\rangle
        )
    $
    \\ \hline
    ${}^3{A}_{2}$ & 
    $
        |e_{y\uparrow}e_{x\uparrow}a_{1\uparrow}a_{1\downarrow}\rangle, ~~
        |e_{y\downarrow}e_{x\downarrow}a_{1\uparrow}a_{1\downarrow}\rangle, ~~
        0.71 (
            |e_{y\uparrow}e_{x\downarrow}a_{1\uparrow}a_{1\downarrow}\rangle
            +
            |e_{y\downarrow}e_{x\uparrow}a_{1\uparrow}a_{1\downarrow}\rangle
        )
    $
    \\
    \hline
\end{tabular}
\label{table:wave_function_nv}
\end{table*}

\subsubsection{Complex defects in w-AlN}
\begin{figure}
    \centering
    \includegraphics[width=0.45 \textwidth]{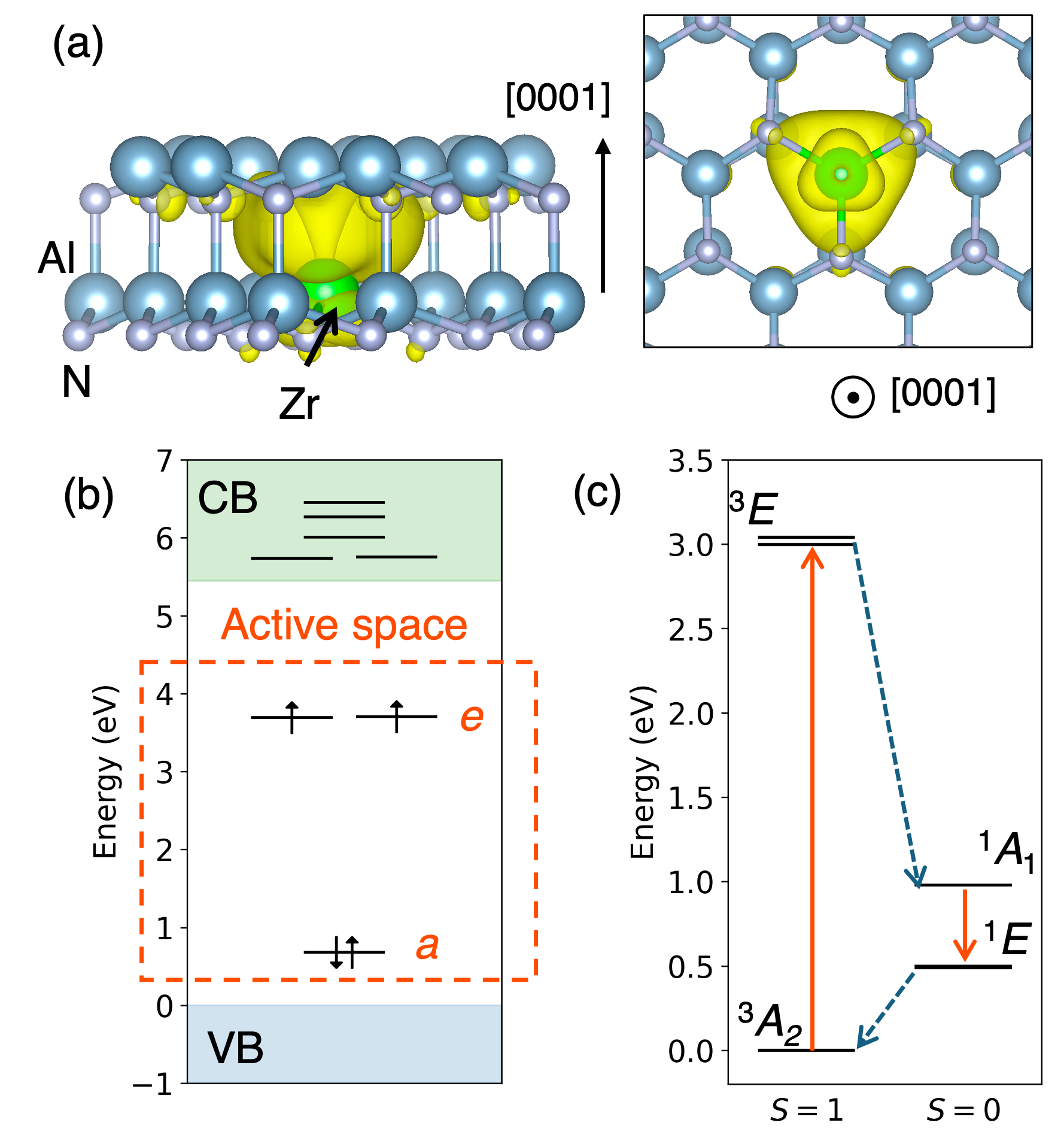} 
\caption{
Same as Figure \ref{fig:dft_nv_diamond} for $(\mathrm{Zr_{Al}V_{N}})^{0}$ in w-AlN.
}
\label{fig:dft_zrv_aln}
\end{figure}

Previous studies \cite{Varley2016PRB, Seo2017PRM} calculated the defect formation energy of charged point defects in a crystal of w-AlN using the finite-size correction of Freysoldt, Neugebauer, and Van de Walle (FNV) \cite{Freysoldt2009PRL}. 
They concluded that complex defects, such as $\mathrm{(Zr_{Al}V_{N})}^{0}$, $\mathrm{(Ti_{Al}V_{N})}^{0}$, and $\mathrm{(Hf_{Al}V_{N})}^{0}$ can be produced. The results for the three complex defects are similar to each other. Thus, in the main text, we explain the results based mainly on $\mathrm{(Zr_{Al}V_{N})}^{0}$. See the other two complex defects in the Appendix. \ref{sec:numerical_results_of_other_vacancies}.

Figure \ref{fig:dft_zrv_aln} presents the classical computational results for the $\mathrm{(Zr_{Al}V_{N})}^{0}$ complex in w-AlN.
Fig. \ref{fig:dft_zrv_aln}(a) presents the spin density obtained from spin-unrestricted DFT calculation.
The spin density of the $(\mathrm{Zr_{Al}V_{N}})^0$ complex exhibits dangling bonds of the Zr atom spreading around the nitrogen defect while holding the $C_{3v}$ symmetry.  
The KS defect levels calculated by spin-restricted DFT calculations were located within the band gap, as shown in Fig. \ref{fig:dft_zrv_aln}(b).
The defect level diagram of the $\mathrm{(Zr_{Al}V_{N})}^{0}$ complex exhibits the same structure as the $\mathrm{NV}^{-}$-center in diamond. 
In addition to the $a_1$, $e_x$ and $e_y$ orbitals, $\mathrm{(Zr_{Al}V_{N})}^{0}$ complex in w-AlN also has defect levels arising from Zr atoms that appear near the conduction band minimum (CBM).
The average value of the transfer integral between the Wannier orbitals ($t = 1.01$ eV) is sufficiently small compared with the energy difference from the $e$ orbital to the CBM. Therefore, the effect of the coupling between the $e$ orbitals and CBM is small, and the defect level can be considered isolated within the band gap.  
Therefore, we adopted an active space that included the $a_1$ and $e$ orbitals.

The results of the calculations for the low-energy effective model are shown in Figure \ref{fig:dft_zrv_aln}(c). 
Theoretical calculations predict that the order of the energy levels will be the same as that in the NV case, although the quantitative values will differ.
Unlike the $\mathrm{NV}^{-}$-center in diamond, a small split of around 0.04 and 0.01 eV occurs in the ${}^3 E$ and ${}^1 E$ states, respectively. 
After checking the wavefunction and symmetry, we concluded that this was because of the finite cell size. 
In the 96-atom supercell calculation, the splitting of the $e_x$ and $e_y$ KS defect levels increases, resulting in a larger split in the ${}^3 E$ and ${}^1 E$ states.
Table \ref{table:wave_function_aln} summarizes the many-body wave function states obtained from FCI calculations. 
It is composed of multiple Slater determinants, which indicates that DFT calculations is difficult to describe the singlet states of $\mathrm{(Zr_{Al}V_{N})}^{0}$ complex in w-AlN.
The excited energies for other vacancy complexes are summarized in Appendix \ref{sec:numerical_results_of_other_vacancies}, where the calculation settings are the same as $\mathrm{(Zr_{Al}V_{N})}^{0}$ complex.

\begin{table}
\renewcommand{\arraystretch}{1.5}
\caption{Computed many-body wave function of the ZrV in w-AlN.}
\label{table:wave_function_aln}
\begin{tabular}{|c|c|}
    \hline
    State   & Many-body wave function  \\
    \hline
    ${}^3{E}$ & 
    $
        |e_{y\uparrow}e_{y\downarrow}e_{x\uparrow}a_{1\uparrow}\rangle, ~~
        |e_{y\uparrow}e_{y\downarrow}e_{x\downarrow}a_{1\downarrow}\rangle, ~~
        |e_{y\uparrow}e_{x\uparrow}e_{x\downarrow}a_{1\uparrow}\rangle,
    $
    \\
    &
    $
        |e_{y\downarrow}e_{x\uparrow}e_{x\downarrow}a_{1\downarrow}\rangle, ~~
        0.71(
            |e_{y\uparrow}e_{y\downarrow}e_{x\downarrow}a_{1\uparrow}\rangle
            +
            |e_{y\uparrow}e_{y\downarrow}e_{x\uparrow}a_{1\downarrow}\rangle
        ),
    $
    \\
    & 
    $
        0.71(
            |e_{y\downarrow}e_{x\uparrow}e_{x\downarrow}a_{1\uparrow}\rangle
            +
            |e_{y\uparrow}e_{x\uparrow}e_{x\downarrow}a_{1\downarrow}\rangle
        )
    $
    \\ \hline
    ${}^1 A_{1}$ &
    $
        0.64|e_{x\uparrow}e_{x\downarrow}a_{1\uparrow}a_{1\downarrow}\rangle
        -
        0.77|e_{y\uparrow}e_{y\downarrow}a_{1\uparrow}a_{1\downarrow}\rangle
    $
    \\ \hline
    ${}^1{E}$ & 
    $
        0.71(
            |e_{y\uparrow}e_{x\downarrow}a_{1\uparrow}a_{1\downarrow}\rangle
            -
            |e_{y\downarrow}e_{x\uparrow}a_{1\uparrow}a_{1\downarrow}\rangle
        ),
    $
    \\
    & 
    $
        0.77|e_{x\uparrow}e_{x\downarrow}a_{1\uparrow}a_{1\downarrow}\rangle
        -
        0.64|e_{y\uparrow}e_{y\downarrow}a_{1\uparrow}a_{1\downarrow}\rangle
    $
    \\ \hline
    ${}^3{A}_{2}$ & 
    $
        |e_{y\uparrow}e_{x\uparrow}a_{1\uparrow}a_{1\downarrow}\rangle, ~~
        |e_{y\downarrow}e_{x\downarrow}a_{1\uparrow}a_{1\downarrow}\rangle, 
    $
    \\
    &
    $   
        0.71 (
            |e_{y\uparrow}e_{x\downarrow}a_{1\uparrow}a_{1\downarrow}\rangle
            +
            |e_{y\downarrow}e_{x\uparrow}a_{1\uparrow}a_{1\downarrow}\rangle
        )
    $
    \\
    \hline
\end{tabular}
\end{table}

\subsection{Quantum computation of the spin defects}
\label{sec:results_quantum}
\subsubsection{Setup}
The quantum circuits were constructed using \texttt{qiskit} v1.1.0 \cite{Qiskit}, where the mapping of the fermionic Hamiltonian to spin operators was performed with \texttt{qiskit-nature} v0.7.2. 
We then translated it into the quantum circuit of \texttt{pytket} with \texttt{pytket} v1.28.0 \cite{pytket} and \texttt{pytket-qiskit} v0.53.0, and executed the quantum circuit with \texttt{pytket-quantinuum} v0.33.0.
We used the Quantinuum H1-1 trapped-ion computer and its emulator (H1-1E) \cite{Ryan2022arXiv, Self2024NPhys}.
The H1-1 quantum computer is composed of 20 qubits and can perform MS gate operations combined with all-to-all connectivity of $\sim 2.0\times 10^{-3}$ infidelity \cite{H1_1}.
State preparation and measurement (SPAM) error is about $3.0\times 10^{-3}$.
We ran 1000 shots of quantum computation.
In Figs. \ref{fig:result_zrv_aln_gs} and \ref{fig:result_zrv_aln_es}, the results of the actual demonstrations on  hardware are represented by H1-1, whereas the other results are based on numerical simulations.

In this study, we calculated the ground and excited states of the singlet state of $\mathrm{(Zr_{Al}V_{N})^{0}}$ in w-AlN.
This state is composed of multiple Slater determinants, as shown in Table \ref{table:wave_function_aln}; therefore, the DFT calculation fails to describe the ground and excited states of the spin singlet state.
When performing quantum calculations, we ignored terms whose absolute value of the Hamiltonian coefficient was less than 0.01.
This approximation resulted in finding the $\mathbb{Z}_2$ symmetry for the $P_{Z_2} = ZIZI$ operator, that is, the Hamiltonian is commutative with the $P_{Z_2}$ operator \cite{Bravyi2017arXiv}.
This symmetry allows for the reduction of the number of qubits by one, and the total number of logical qubits, including an ancilla qubit of PITE, becomes four, which can be implemented in the quantum circuit in Figure \ref{fig:circuit_pite}.
In addition, the effect of the approximation is not observed at the second decimal point.

By running our circuit $N_{\mathrm{shot}}$ times on quantum simulator or computer, we evaluate  mean estimator $\bar{O}_{\rho}$ by measuring $O$ with respect to quantum state $\rho$ \cite{Cai2023RMP}.  The deviation of mean estimator $\bar{O}_{\rho}$ from the true value $\mathrm{Tr}[O\rho_0]$ obtained from the noiseless state $\rho_0$ is assessed by the mean squared error (MSE) given by  \begin{gather}
    \mathrm{MSE}[\bar{O}_{\rho}]     
    =     
    \mathrm{bias}[\bar{O}_{\rho}]^2     
    +     
    \mathrm{var}[\bar{O}_{\rho}], 
\label{eq:mse} 
\end{gather} 
where the bias and the variance is calculated as \begin{gather}     
    \mathrm{bias}[\bar{O}_{\rho}]     
    =     
    \mathrm{Tr}[O\rho]     
    -     
    \mathrm{Tr}[O\rho_0]     
    ,     
    \nonumber \\     
    \mathrm{var}[\bar{O}_{\rho}]     
    =     
    \frac{         
        \mathrm{Tr}[O^2\rho]         
        -         
        \mathrm{Tr}[O\rho]^2     
    }{N_{\mathrm{cir}}},  
    \label{eq:bias_var} 
\end{gather} 
with the number of available outcomes, $N_{\mathrm{cir}}$. Bias and variance are often called systematic and statistical errors, respectively. In this study, we adopted projection operators as observable $O$.

\subsubsection{Ground state of the singlet state}

\begin{figure*}
    \centering
    \includegraphics[width=0.9 \textwidth]{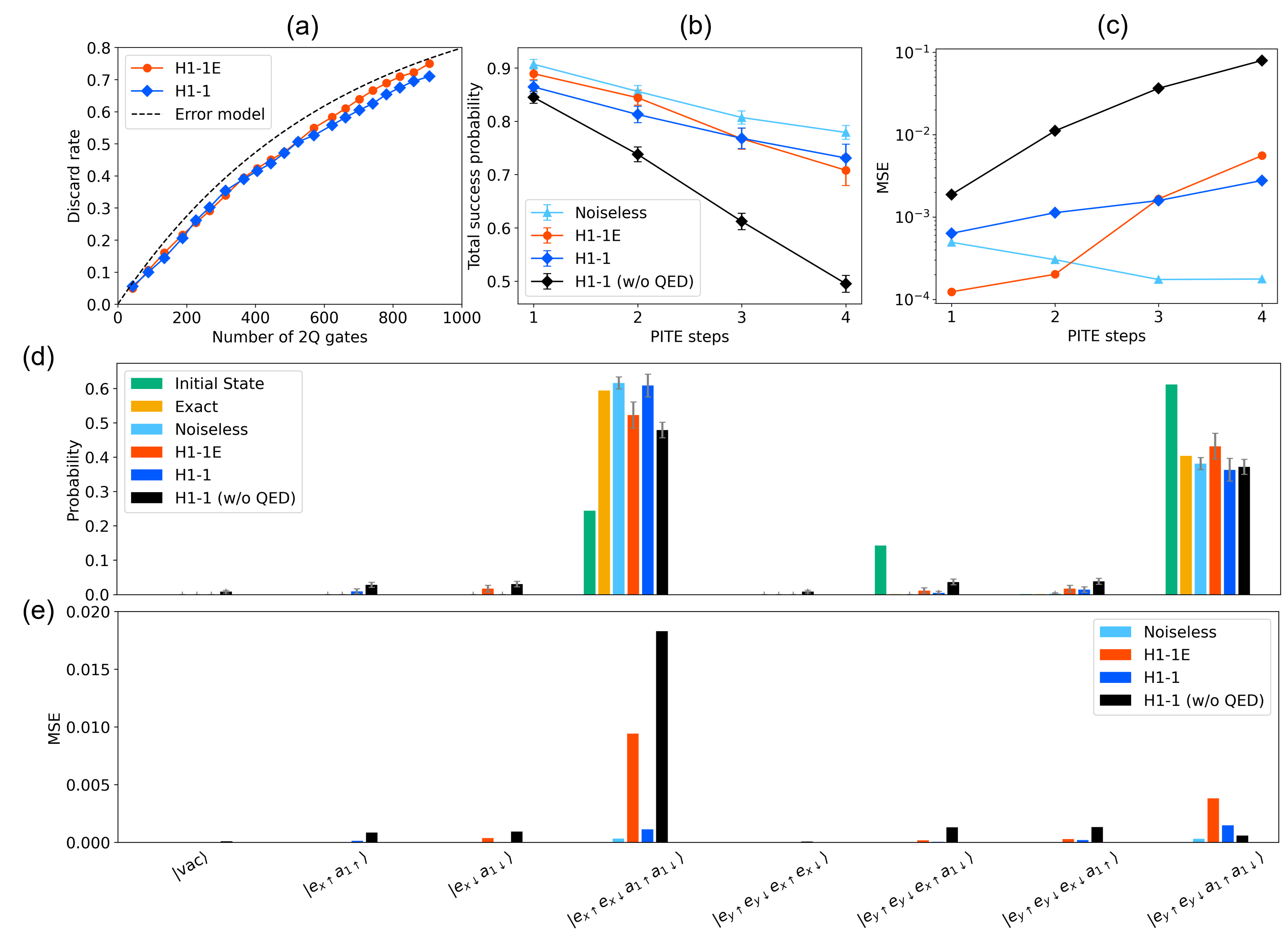} 
\caption{
Computation results for $\mathrm{(Zr_{Al}V_{N})^{0}}$ complex in w-AlN obtained using a quantum computer.
(a) Discard rates of the calculated results owing to errors detected by QED as a function of the number of two-qubit gates. The dashed line denotes the noise model shown in Eq. (\ref{eq:noise_fit}), with $p_2 = 1.6 \times 10^{-3}$. 
(b) Total success probability of PITE and (c) its MSE according to the PITE steps. 
(d) Histogram of measured computational basis and (e) its MSE. 
Noiseless denotes the noiseless simulation of the PITE, and H1-1E denotes the emulator of the H1-1 quantum computer. The results denoted by H1-1 are obtained from actual demonstrations on hardware, whereas the others are based on numerical simulations.
The error bars in Figs. (b) and (d) denotes the statistical error.
All calculations used $N_{\mathrm{shot}} = 1000$. 
}
\label{fig:result_zrv_aln_gs}
\end{figure*}

\begin{figure*}
    \centering
    \includegraphics[width=0.85 \textwidth]{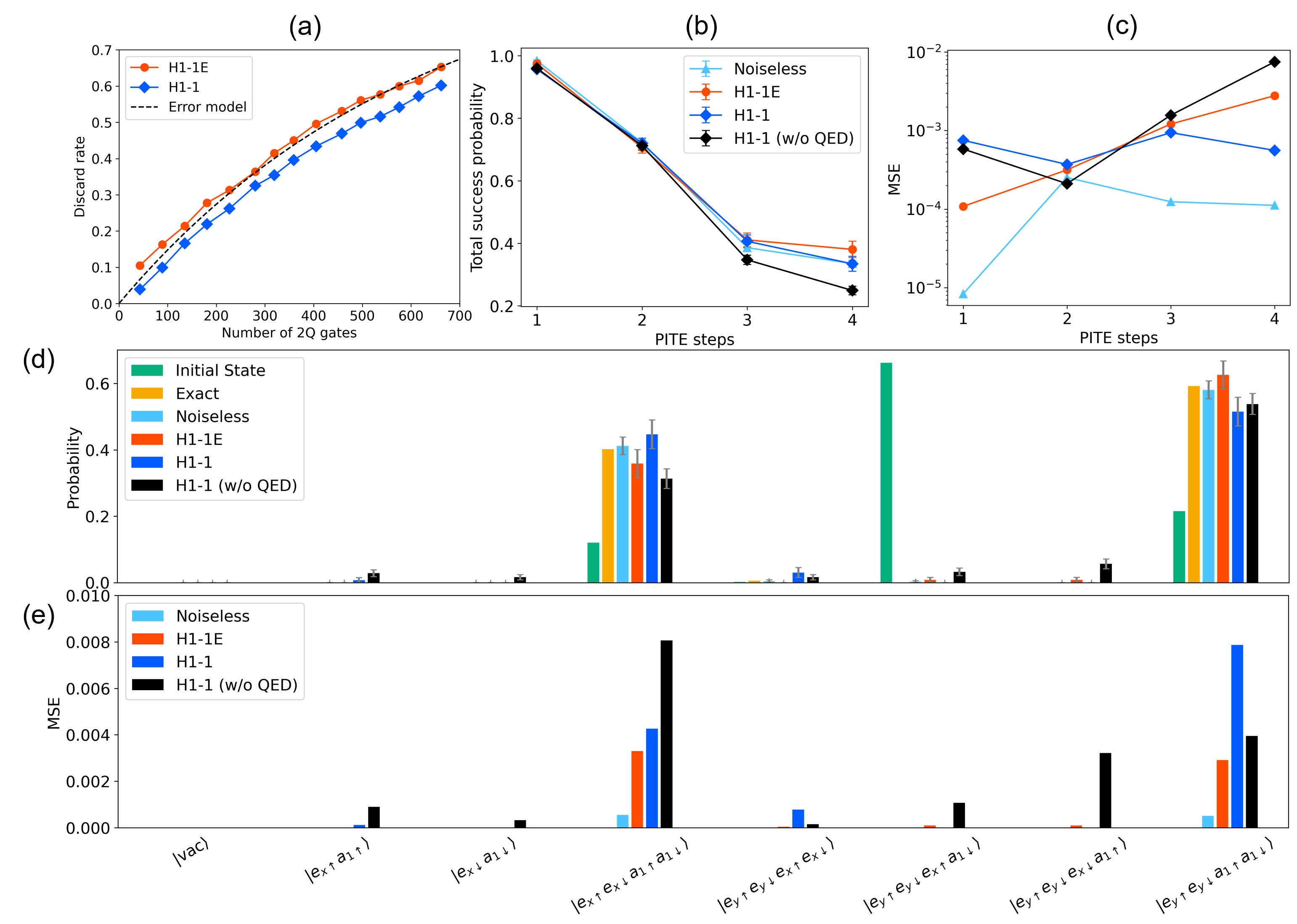} 
\caption{
Same as Figure \ref{fig:result_zrv_aln_gs} for the excited state.
}
\label{fig:result_zrv_aln_es}
\end{figure*}

In this study, we prepared a good initial state for PITE as
$
    |\psi\rangle
    =
    \left(
    3|\phi_{\mathrm{gs}}\rangle
    +
    |\phi_{\mathrm{es1}}\rangle
    +
    |\phi_{\mathrm{es2}}\rangle
    +
    |\phi_{\mathrm{es3}}\rangle
    \right) / \sqrt{12}
$, owing to hardware limitation.
The drawback of the original PITE method is that the total success probability exponentially decreases with respect to $m_0$ as the number of PITE steps increases \cite{Nishi2023PRRes}.
To avoid this drawback, a constant energy shift was used \cite{Nishi2023PRRes}.
Note that a constant energy shift requires ground-state energy, and the approximated ITE operator in Eq. (\ref{eq:approximated_pite}) will be the cosine operators, $\cos(\Delta E_k \Delta\tau s_1)$, where the approximated PITE circuit is related to cosine filtering \cite{Ge2019JMP} and Rodeo's algorithm \cite{Choi2021PRL} except for parameter choice. 
In a constant energy shift, the total success probability becomes $\mathbb{P}_0 = |c_1|^2$.
The details of the computational cost of the PITE method were analyzed in \cite{Nishi2023PRRes}.
We performed four steps of the PITE operations.
We linearly changed $\Delta \tau_{k}$ from 0.25 to 
$
    \pi / [2 s_1 (E_{\mathrm{es1}} - E_{\mathrm{gs}})]
    = 2.41
$, where $m_0 = 0.8$.
Because the accuracy of the Trotter-Suzuki decomposition for CRTE gates worsens as $\Delta \tau_{k}$ increases, the dividing numbers for $\Delta \tau_{k}$ are chosen as $r = (1, 2, 3, 4)$.
At each step of the PITE, the number of syndrome measurements performed  is (2, 4, 6, 8), and the number of two-qubit gate operations, including the syndrome measurement, is (92, 178, 257, 336). 
A total of 21 syndrome measurements were performed.

Figure \ref{fig:result_zrv_aln_gs}(a) shows the results of the discard rate obtained by the QED.
In this study, we employed the QED code to protect against quantum noise, which necessitated discarding a portion of the measurement outcomes.
The quantum error of H1-1 quantum computer is modeled as depolarizing error and a fitting function of the discard rate is given by \cite{Yamamoto2024PRR}
\begin{gather}
    d(N_{\mathrm{2Q}}, p_2)
    =
    1 - (1 - p_2)^{N_{\mathrm{2Q}}} ,
\label{eq:noise_fit}
\end{gather}
where $N_{\mathrm{2Q}}$ denotes the number of two-quantum gates within the circuit.
We adopted the fitting parameter $p_2 = 1.6 \times 10^{-3}$ used in a previous demonstration \cite{Yamamoto2024PRR} and confirmed that it also agrees well with our demonstrations.
It is clearly seen that the discard rates of H1-1 and H1-1E agree well with each other, confirming that the quantum noise that occurs on the actual device can be well modeled by the emulator.
We mention the reasons behind the relatively moderate discard rate of 71\%. 
The PITE circuit contains the several multi-qubit Pauli rotation gates, which are typically decomposed using a combination of several CNOT and single-qubit gates. In the case of four data qubits, the three- and four-qubit rotation gates can be implemented as a single MS gate as given by Eqs. (\ref{eq:logical_pauli_x}), (\ref{eq:logical_pauli_z}), and (\ref{eq:logical_pauli_y}). Therefore, the number of two-qubit gates was not excessive, resulting in a moderate value of the discard rate.
We also remark on the applicability of the noise model in Eq. (\ref{eq:noise_fit}) to other hardware platforms. In Refs. \cite{Google2023Nature, Google2024Nature}, the authors investigated the detection correlation matrix using surface code and compared it with a noise model simulation and that obtained using actual hardware composed of transmon qubits. The noise model based on the depolarizing error underpredicts the results, and the inclusion of leakage, crosstalk, and stray interactions improves the results and achieves a result compatible with the experimental result.

% -- total success probability
The total success probability for the PITE steps is presented in Figs. \ref{fig:result_zrv_aln_gs}(b).
 The total success probability of the noiseless result monotonically decreases, approaching the overlap between the initial and ground states.  Note that the true value $\mathrm{Tr}[O\rho_0]$, which is employed by the Trotter-Suzuki decomposition, is not the same as the exact value that excludes the algorithmic error. In this calculation, the noiseless result gives a value of 0.78, which is slightly larger than the exact value of 0.75.   Figure \ref{fig:result_zrv_aln_gs}(c) shows the MSE of the total success probability, where the MSE of the noiseless simulation is relatively low in most cases. This reason is that the MSE of the noiseless simulation is primarily influenced by statistical error.  
 % ---- H1-1 w/o QED 
 Considering quantum noise, a reduction in the total success probability from the noiseless one was observed, as presented in the results without QED code [Fig. \ref{fig:result_zrv_aln_gs}(b)].  The MSE of the results without QED codes monotonically increases, as shown in Fig. \ref{fig:result_zrv_aln_gs}(c).  This is due to the fact that the systematic error based on quantum noise increases in accordance with the PITE step.  Note that when $N_\mathrm{shot}$ is fixed, the statistical error is reduced in accordance with the reduction in the total success probability from one, as illustrated in Eq. (\ref{eq:bias_var}) ( the noiseless MSE in Fig. \ref{fig:result_zrv_aln_gs}(c)).  
 % ---- with QED 
 When employing QED code, the total success probabilities become closer to the noiseless result.  The statistical error of the results without QED code does not change through the PITE step proceeding, whereas the QED-employed cases increase as the PITE step increases, as shown by the error bars in Fig. \ref{fig:result_zrv_aln_gs}(b).  The reason for the increase in the statistical error is clear from the reduction in $N_{\mathrm{cir}}$ due to the increase in the discard rate, which is calculated as $N_{\mathrm{cir}} =  [1-d(N_{\mathrm{2Q}}, p_2)]N_{\mathrm{shot}}$. 
 % ---- summary 
 That is, the QED code has a trade-off of MSE between the reduction of the systematic error and the increase in the statistical error, as given by Eq. (\ref{eq:mse}). In Fig. \ref{fig:result_zrv_aln_gs}(c), the MSE for H1-1E and H1-1, when employing the QED code, exhibits a relatively low value of MSE in comparison to the MSE of H1-1 without QED code.  This indicates that the benefit of reducing system errors is greater than the cost of increasing statistical errors. We found that the iceberg code effectively mitigates the impact of quantum noise on both numerical simulations and actual demonstrations on hardware.  To further enhance the precision of the results, an increase in the number of measurements $N_{\mathrm{shot}}$ allows for an improvement in the accuracy of the results in proportion to $O(1/\sqrt{N_{\mathrm{shot}}})$.

% -- Slater determinants
The measured probabilities of each Slater determinant, that is, the absolute square of the coefficients of the Slater determinant, are shown in Figure \ref{fig:result_zrv_aln_gs}(d, e).
There is a slight deviation between the results of the FCI calculation and the noiseless simulation of the PITE for the effective Hamiltonian. 
The deviation between the exact and noiseless values is attributed to two sources: an algorithmic error based on the Trotter-Suzuki decomposition, and a statistical error resulting from the finite number of measurements.
In the histogram in Figure \ref{fig:result_zrv_aln_gs}(d), H1-1 and H1-1E are in good agreement, as in the case of total success probability. 
In both results, a slight amplitude was observed in states other than $|e_{x\uparrow} e_{x\downarrow} a_{1\uparrow} a_{1\downarrow}\rangle$ and $|e_{y\uparrow} e_{y\downarrow} a_{1\uparrow} a_{1\downarrow}\rangle$.
This is considered to be due to the influence of quantum errors that are not detected by the QED, such as two-qubit errors. 
 Furthermore, the MSE for the measured probabilities of each Slater determinant was evaluated, as illustrated in Fig. \ref{fig:result_zrv_aln_gs}(e). In most cases, the MSE for H1-1 without QED exhibited considerable MSEs. This illustrates the efficacy of the iceberg code.
We quantify the results using classical fidelity \cite{Nielsen2000Book} defined as
\begin{gather}
    F(|\Psi^{\mathrm{FCI}}\rangle, |\Psi\rangle) 
    :=
    \left(
        \sum_{i} |c_i^{(\mathrm{FCI})}|  |c_i^{\beta}| 
    \right)^2,
\end{gather}
where $|\Psi^{\mathrm{FCI}}\rangle = \sum_{i} c_i |i\rangle$ represents the ground state obtained by FCI calculation and $|\Psi\rangle$ is ground state by PITE.
The classical fidelity of H1-1 with QED is 0.98, and we can observe that the results have improved compared with the classical fidelity of 0.87, which does not use QED.  
We note that the unencoded circuit contains 743 two-qubit gates.
As a result, this demonstration indicates that the influence of quantum errors that are not detected by the QED is small, and the ground state of the singlet state is successfully obtained in a trapped-ion quantum computer.  
The calculation of physical quantities is a crucial aspect of this study. Although the calculation of energy is useful for assessing the precision of the calculation, it necessitates a deeper circuit when employing QPE \cite{Yamamoto2024PRR} or conducting multiple measurements with a change in Pauli basis. In order to calculate the fidelity, it is necessary to apply state tomography.  However, both QPE and state tomography must be applied to logical states, which may result in a significant increase in computational cost. In this study, the inability to provide energy was constrained by the limitations of hardware resources.  However, the results and discussion, including the physical quantities, will be possible in the future.

The calculation results using the emulator for Quantinuum H1-1 for $\mathrm{NV}^{-}$ center in diamond, and $\mathrm{(Hf_{Al}V_{N})^{0}}$ and $\mathrm{(Ti_{Al}V_{N})^{0}}$ complexes in w-AlN are summarized in Appendix \ref{sec:numerical_results_of_other_vacancies}.

\subsubsection{Excited state of spin singlet}
Using a filter circuit \cite{Meister2022arXiv, Kosugi2023JJAP} which removes states with a given eigenvalue, the excited state can be obtained.
However, a filter circuit must also be implemented in addition to the quantum circuit used to obtain the ground state; therefore, it is difficult to execute the excited-state calculation on current quantum hardware limited by the number of gate operations. 
In this study, we assume that the ground state has been obtained using techniques such as quantum tomography, and we use the initial state after removing the ground state.

In the calculation, the initial state is prepared as 
$
    |\psi\rangle
    =
    \left(
    |\phi_{\mathrm{es1}}\rangle
    +
    |\phi_{\mathrm{es2}}\rangle
    +
    |\phi_{\mathrm{es3}}\rangle
    \right) / \sqrt{3}
$.
We performed four steps of the PITE operations.
We linearly changed $\Delta \tau_{k}$ from 0.05 to 
$
    \pi / [2 s_1 (E_{\mathrm{es2}} - E_{\mathrm{es1}})]
    = 0.57
$, where $m_0 = 0.8$.
As $\max \Delta \tau_k$ is smaller than that of the ground-state calculation, the dividing numbers for $\Delta \tau_{k}$ are $r = (1, 1, 2, 3)$.
Fifteen syndrome measurements were used for the calculations.
The encoded and unencoded quantum circuits contained 662 and 537 two-qubit gate operations, respectively.

As with the ground-state calculation, the discard rate is shown in Fig. \ref{fig:result_zrv_aln_es}(a). 
The discard rate for H1-1 is 0.60, and that for H1-1E is 0.65.
H1-1 has slightly less quantum noise than H1-1E; however, both are consistent with the fitting of the noise model in Eq. (\ref{eq:noise_fit}).

The  total success probability for the PITE is shown in Fig. \ref{fig:result_zrv_aln_es}(b). The results of the noiseless simulation and encoded simulation (H1-1 and H1-1E) show similar values, but it was observed that the total success probability is smaller in the case without QED.  
 The MSE is plotted in Fig. \ref{fig:result_zrv_aln_es}(c). Reflecting the small number of two qubit gates included in the circuit, the MSE for the result without QED is smaller than that for the ground state.  The utilization of QED results in a reduction in MSE compared to instances where QED is not employed.  The effectiveness of QED in the calculation of excited states was also observed.

The measured probability of Slater determinants and is MSE are presented in Fig. \ref{fig:result_zrv_aln_es}(d) and (e), respectively.
It can be confirmed that the classical fidelities of H1-1 and H1-1E using QED are 0.98, and that QED improves on the classical fidelity of 0.87 which does not use QED.

\section{Conclusions}
A low-energy effective model for calculating the electronic state of a quantum spin was constructed, and the probabilistic imaginary-time evolution (PITE) algorithm, designed for the fault-tolerant quantum computer (FTQC) era, was applied to the effective Hamiltonian in order to calculate the ground state and excited states of the singlet state on a Quantinuum trapped-ion quantum computer.
In order to address the quantum noise that occurs in quantum hardware, the PITE method was implemented using the $\llbracket n+2, n, 2 \rrbracket$ quantum error detection (QED) code, referred to as the Iceberg code.
The Quantinuum hardware facilitates mid-circuit measurements and reuse, which is advantageous for QED syndrome measurements and ancilla-qubit measurements in PITE.
In quantum computation with an encoded state, it is essential to measure the ancilla qubit of PITE at each PITE step. However, this measurement results in the destruction of the encoded state. To address this challenge, we employed re-encoding to recover the encoded state.
Furthermore, we concentrated on the fact that the low-energy effective Hamiltonian can be divided into a large diagonal term resulting from the one-body term and a small non-diagonal term based on the two-body interaction. We also developed a circuit implementation that reduces the number of two-qubit gates based on second-order Trotter-Suzuki decomposition.

An effective low-energy model has been constructed for the NV centre in diamond, which is a promising quantum sensing material, and for complex defects in w-AlN, which is expected to be a new spin-qubit material.
The constructed effective Hamiltonian was confirmed to correctly describe the energy order of the excitation energies.
We performed calculations of the ground and excited states of $\mathrm{Zr_{Al} V_N}$ complex in w-AlN using the Quantinuum H1-1 quantum computer.
The quantum circuit for PITE with QED included up to 906 two-qubit gates, with a discard rate of 71\% due to the QED. The implementation of QED confirmed that the ground and excited states can be obtained with a classical fidelity of 0.98.

In this study, the simulation results were discarded when quantum errors were detected using QED. However, this resulted in an increase in the variance of finite measurements. Therefore, it would be beneficial to implement quantum error correction.
We were unable to display the energy and fidelity in this study, which is an important quantity in the field of quantum chemistry, owing to the limited hardware resources. 
One calculation method with a constant depth overhad is to use quantum-state tomography, which causes scalability problems.   
Phase estimation enables the provision of energy with an additional circuit depth and avoids an increase in the number of measurements, which will be available in future protocols.
In addition, since we successfully prepared the ground state, we will calculate such physical quantities, including one-body Green's function and response function \cite{Kosugi2020PRA, Kosugi2020PRRes}. 
Furthermore, due to the constraints of the hardware, the minimum active space, comprising the $a$ and $e$ orbitals, was selected for the calculations. However, it would be more optimal to utilize a broader active space to enhance the precision of the calculations.
As the number of orbitals included in the active space increases, the computational complexity rises exponentially, rendering classical computers inadequate for such calculations. Consequently, the potential of quantum computers is expected to expand considerably \cite{Yoshioka2024npjQI}.
As PITE has already been demonstrated to be quadratically accelerated \cite{Nishi2024PRR}, it may be possible in the future to provide experimental evidence that quantum computation is more effective than classical computation in practical calculations based on advances in hardware.

\section*{ACKNOWLEDGMENTS}
The author acknowledges the contributions and discussions provided by Prof. Shinji Tsuneyuki and the members of Quemix Inc.
% supercomputer
The authors thank the Supercomputer Center, the Institute for Solid State Physics, the University of Tokyo for the use of the facilities.
% Kakenhi
This work was supported by JSPS KAKENHI under Grantin-Aid for Scientific Research No.21H04553, No.20H00340, and No.22H01517, JSPS KAKENHI. This work was partially supported by the Center of Innovations for Sustainable Quantum AI (JST Grant Number JPMJPF2221).

\appendix
\section{Computational details of the DFT calculation}
\label{appendix:dft_calculation}
\subsection{Lattice parameters}
The lattice constant of diamond was calculated as $a = 3.534$ \AA, which was obtained from spin-restricted DFT calculations with PBE functional. 
The calculated lattice constant is in good agreement with the experimental value of $a = 3.567$ \AA \cite{Madelung2012}. 
The calculated band gap is $E_g = 4.19$ eV, which underestimates the experimental value of $E_g$ = 5.45--5.50 eV.
This underestimation can be improved to $E_g = 5.39$ eV by using the screened hybrid functional of Heyd, Scuseria, and
Ernzerhof (HSE) \cite{Heyd2003JCP, Heyd2006JCP}, starting from a structure with a lattice constant of $a = 3.534$ \AA.

We used the previously reported structural properties of w-AlN: $a = 3.130$ \AA, $c/a = 1.603$, and $u=0.382$ \cite{Seo2016SciRep}. 
The experimental values were $a = 3.110$ \AA, $c/a = 1.601$, and $u=0.382$ \cite{Rinke2008PRB}.
However, similar to diamond, the band gap is underestimated by our calculation with the PBE function, as $E_g = 4.09$ eV.
This underestimation was also improved by the HSE function, as in $E_g = 5.36$ eV. 
The lattice parameters computed in the DFT calculation with the PBE function are used in the supercell calculation in the main text.

\subsection{Convergence of the system size of w-AlN}
\begin{figure*}[ht]
    \centering
    \includegraphics[width=0.95 \textwidth]{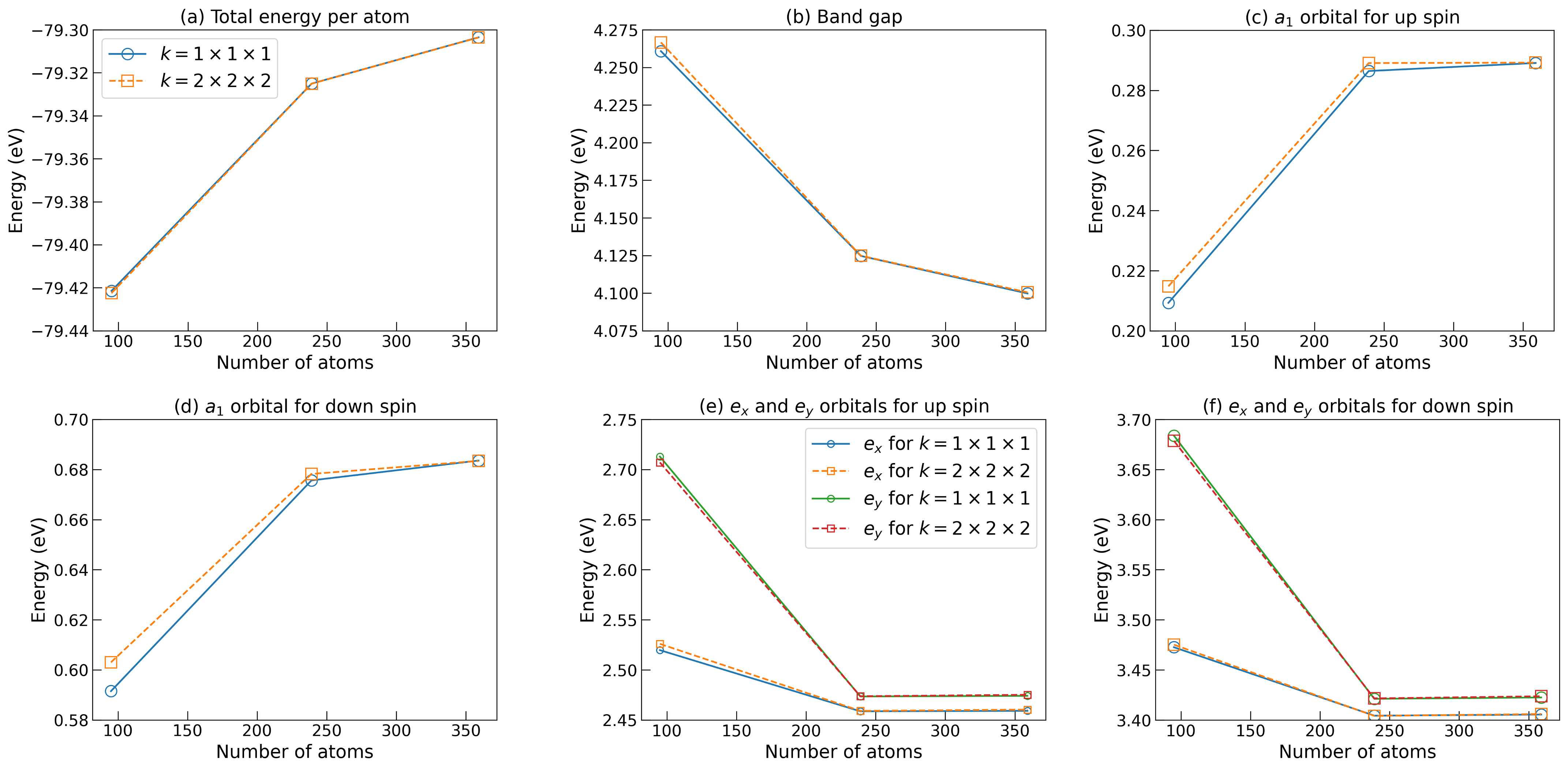} 
\caption{
Numerical errors of the defect calculation for $\mathrm{(Zr_{Al}V_{N})}^{0}$ complex in w-AlN introduced by the supercell size and $k$-ponit sampling. 
(a) Total energy per atom, (b) band gap, and (c)–(f) single-particle states calculated using the spin-unrestricted KS equation as a function of supercell size and $k$-point sampling. 
96-, 240-, and 360-atoms supercell were used.
All points were calculated with the PBE functional and plane-wave cutoff of 75 Ry.
}
\label{fig:convergence_supercell_wAlN_ZrV}
\end{figure*}

The convergence of supercell size was examined by varying the number of unit cells in the supercell. Supercells containing 96, 240, or 360 atoms were used.
The $k$-point sampling convergence was studied using Monkhorst-Pack grids. The densities of $k$-points were taken as $1\times 1\times 1$ and $2 \times 2 \times 2$.
The total energy, band gap, and single-particle states in the band gap are shown in Fig. \ref{fig:convergence_supercell_wAlN_ZrV}.

Similar to a previous investigation of the numerical error in the negatively nitrogen vacancy ($\mathrm{V_{N}^{-1}}$) in w-AlN \cite{Seo2016SciRep}, we also observed that $\Gamma$-point sampling is not sufficient in a 96-atoms supercell. 
However, a slight difference in the quantities between $\Gamma$ and $2\times 2\times 2$ $k$-point sampling was observed in the 240-atoms supercell. 
In Fig. \ref{fig:convergence_supercell_wAlN_ZrV}(b), the band gap decreases and converges to 4.10 eV, which is close to the value of perfect crystal of w-AlN, $E_g=4.09$ eV.  
Although single-particle states $e_x$ and $e_y$ have a small splitting in the 96-atoms supercell calculation in Fig. \ref{fig:convergence_supercell_wAlN_ZrV} (e) and (f), these two states close each other and almost degenerate at 240-atom supercell. 
Accordingly, we choose a 240-atoms supercell and $\Gamma$ point sampling in the main text.

\subsection{Convergence of constructing the low-energy effective Hamiltonian}
We investigated the convergence of the vertical excitation energy for the NV$^{-}$-center in diamond as a function of the parameters contained in the RESPACK software \cite{Nakamura2021CPC} when an effective low-energy model was constructed.
The KS wave function as an input for the RESPACK calculation was obtained by spin-restricted DFT calculation with the PBE functional starting from the optimized structure of the 216-atom supercell.
Note that we increased the energy cutoff of the plane-wave basis sets from 50 to 100 Ry to properly describe the high-energy states. 
An insufficient energy cutoff induces symmetry breaking of the effective Hamiltonian, for example,. for all $(i, j)$ pairs, where $i\neq j$ and Coulomb integrals $V_{ij}$ are the same, whereas screened Coulomb integrals $U_{ij}$ are different from each other. 
For example, in this case, the degeneracy of ${}^1E$ state is broken.

\begin{figure}[ht]
    \centering
    \includegraphics[width=0.4 \textwidth]{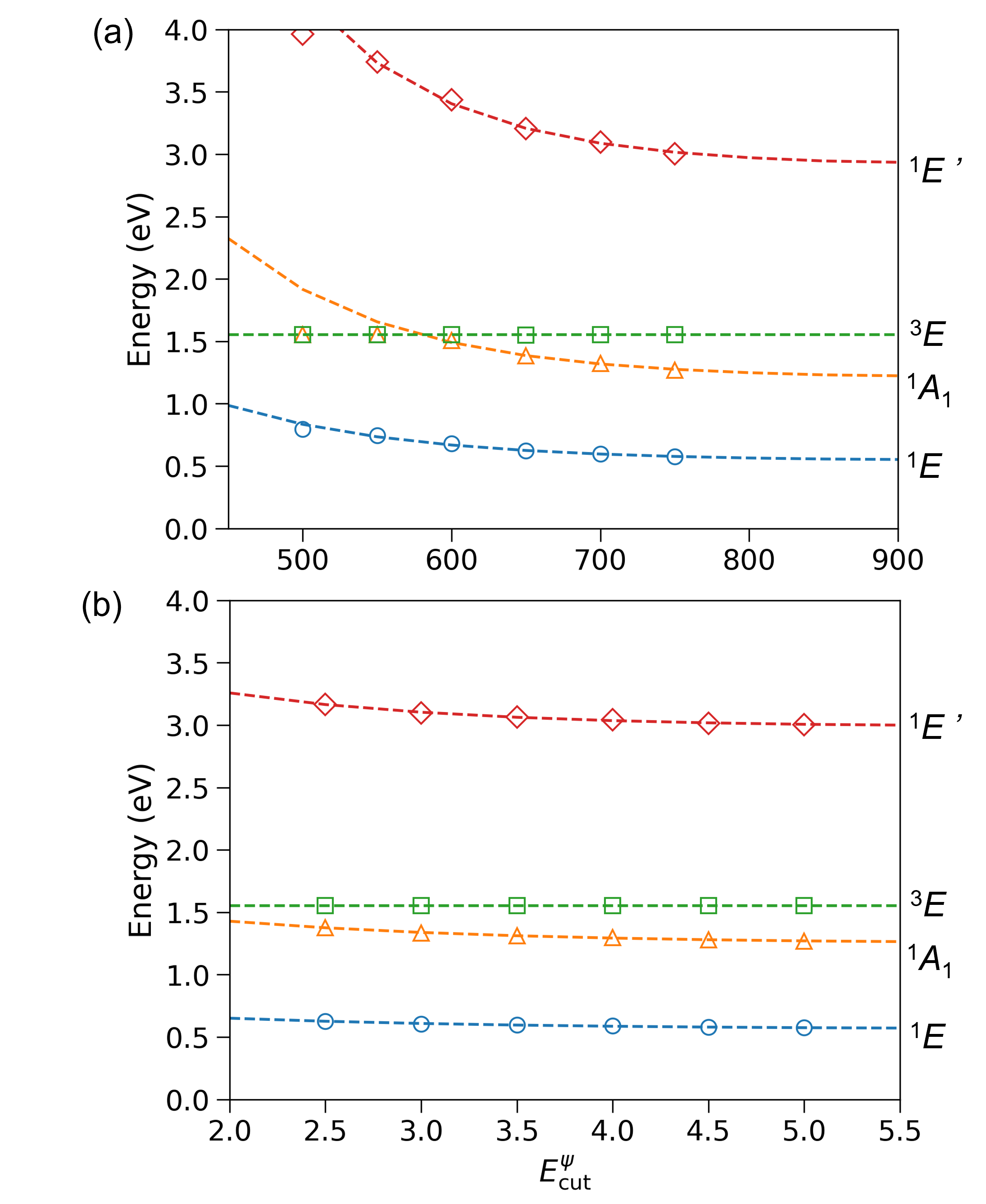} 
\caption{
FCI energies for the NV$^{-}$-center in diamond as a function of (a) number of bands, $N_{\mathrm{band}}$, and (b) cutoff energy for the dielectric function, $E_{\mathrm{cut}}^{\varepsilon}$. 
A 216-atom supercell and $\Gamma$point sampling were used.
All points were calculated with a PBE functional and plane-wave cutoff of 100 Ry.
}
\label{fig:fig_convergence_respack}
\end{figure}

We varied two parameters: the number of bands, $N_{\mathrm{band}}$ and the cutoff energy for the dielectric function, $E_{\mathrm{cut}}^{\varepsilon}$. 
The FCI energy is fitted by \cite{Yoshida2024JCP}
\begin{gather}
    f(x) = b \times \exp(-x/c) + \Delta E_{\infty}
\label{eq:respack_fit}
\end{gather}
where $b$ and $c$ are real numbers and $\Delta E_{\infty}$ denotes the extrapolated excitation energy. 
We consider $x$ as the number of bands $N_{\mathrm{band}}$ in Fig. \ref{fig:fig_convergence_respack}(a) and the cutoff energy for dielectric function $E_{\mathrm{cut}}^{\varepsilon}$ in Fig. \ref{fig:fig_convergence_respack}(b).

The dependence of the FCI energy on $N_{\mathrm{band}}$ is plotted in Fig. \ref{fig:fig_convergence_respack}(a), where $E_{\mathrm{cut}}^{\varepsilon}$ is fixed to 5.0 eV.
In a small number of bands, such as $N_{\mathrm{band}}$ = 500 -- 550, ${}^3E$ and ${}^1A_1$ states are almost degenerate because of the lack of inclusion of high-energy bands. 
Increasing $N_{\mathrm{band}}$ decreases the excitation energy. 
and we almost reach convergence at $N_{\mathrm{band}} = 750$.
We fitted equation (\ref{eq:respack_fit}) at three points: 650, 700, and 750.
The results obtained at $N_{\mathrm{band}}=750$ (0.57/1.27/1.55 eV for ${}^1E/{}^{1}A_{1}/{}^{3}E$ states) are different from the extrapolated results $\Delta E_{\infty}$ (0.54/1.20/1.55 eV for ${}^1E/{}^{1}A_{1}/{}^{3}E$ states) within 0.07 eV.

Figure \ref{fig:fig_convergence_respack}(b) shows the dependence of excitation energy on $E_{\mathrm{cut}}^{\varepsilon}$, where $N_{\mathrm{band}}=750$.  Unlike Fig. \ref{fig:fig_convergence_respack}(a), the excitation energy does not vary much with respect to $E_{\mathrm{cut}}^{\varepsilon}$.
The results obtained at $E_{\mathrm{cut}}^{\varepsilon} =$ 3.0 Ry (0.61/1.33/1.55 eV for ${}^1E/{}^{1}A_{1}/{}^{3}E$ states) is sufficiently converged to the extrapolated results (0.56/1.25/1.55 eV for ${}^1E/{}^{1}A_{1}/{}^{3}E$ states).
It is known that $E_{\mathrm{cut}}^{\varepsilon}$ should be 0.1 times the kinetic energy cut-off used in the DFT calculation \cite{Nakamura2021CPC} however, the results in Fig. \ref{fig:fig_convergence_respack}(b) show that a reduction to 0.06 times is valid.

\section{Comparison of vertical excitation energy with other methods}
\label{sec:comparison_other_method}

\begin{table*}[ht]
    \centering
    \renewcommand{\arraystretch}{1.3} % Adjust row height
    \setlength{\tabcolsep}{6pt} % Adjust column spacing
    \newcolumntype{L}{>{\raggedright\arraybackslash}X}
    \newcolumntype{Y}{>{\centering\arraybackslash}X}
    \caption{Computed vertical excitation energies (eV) of the negatively charged NV$^{-}$-center in diamond.}
    \begin{tabular}{ | m{5.5cm} | m{2cm} m{2cm} m{2cm} m{2cm} m{2cm}| } 
    % \begin{tabularx}{\columnwidth}{|L|YYYYY|}
        \hline
        Method & ${}^3A_2 \to {}^3E$ & ${}^1E   \to {}^1A_1$ & ${}^1A_1 \to {}^3E$ 
        & ${}^3A_2 \to {}^1A_1$ & ${}^3A_2 \to {}^1E$ \\
        \hline
        GW+BSE \cite{Ma2010PRB}    
        & 2.09 & 0.59 & 1.10 & 0.4 & 0.99 \\
        CI on $\mathrm{C_{42}H_{42}N}$ \cite{Delaney2010Nano}
        & 1.93 & 1.43 & -0.1 & 1.94 & 0.51 \\
        Ext. Hubb. + DFT par. \cite{Ranjbar2011PRB}
        & 2.38 & 0.62 & 1.35 & 0.41 & 1.03 \\
        Ext. Hubb. + GW fit. \cite{Choi2012PRB}
        & 2.0  & 0.96 & 0.6 & 1.4 & 0.44\\
        CI + cRPA \cite{Bockstedte2018npjQM} 
        & 2.05 & 0.89 & 0.69 & 1.36 & 0.47 \\
        CI + RPA \cite{Ma2020npjCM}   
        & 1.92 & 0.9  & 0.55 & 1.37 & 0.47 \\
        CI + beyond-RPA \cite{Ma2020npjCM} 
        & 2.00 & 1.2 & 0.24 & 1.76 & 0.56 \\
        QDET \cite{Sheng2022JCTC}
        & 2.15 & 0.81 & 0.88 & 1.27 & 0.46 \\
        TDDFT (DDH) \cite{Jin2023ACS}    
        & 2.11 & 1.4 & 0.15 & 1.96 & 0.56 \\
        This study (PBE + Ext. Hubb + CI) 
        & 1.55 & 0.73 & 0.22 & 1.33 & 0.6 \\
        This study (HSE + Ext. Hubb + CI) 
        & 2.12 & 0.82 & 0.66 & 1.46 & 0.64 \\
        \hline
        Experiment      
        & 1.945 \cite{Davies1976} 
        & 1.190 \cite{Rogers2008NJP}
        & $\approx$ 0.4 \cite{Thiering2017PRB} & & \\
        \hline
    \end{tabular}
    %\end{tabularx}
    \label{table:previous_excitation_energy_nv_diamond}
\end{table*}

The previous numerical results for the exciton energy for NV center diamonds are summarized in Table \ref{table:previous_excitation_energy_nv_diamond}.
See also the review paper on spin-defect calculations \cite{Ivady2018npjCM}.

\section{Model parameters}
\label{sec:model_paremeters}
\subsection{Model parameters of the low-energy effective Hamiltonian}

\begin{table*}[ht]
    \centering
    \caption{Model parameters of the low-energy effective Hamiltonian for $(\mathrm{Zr_{Al}V_{N}})^{0}$, $(\mathrm{Hf_{Al}V_{N}})^{0}$, $(\mathrm{Ti_{Al}V_{N}})^{0}$ complexes in w-AlN, and NV$^-$-center in diamond. The unit of $t_{ij}$, $U_{ij}$, and $J_{ij}$ is eV. The parameters have symmetry as $(i,j)=(j,i)$.}
    \begin{tabular}{
        | >{\centering}p{2.7cm}
        | >{\centering}p{2.7cm}
        | >{\raggedleft}p{1.8cm}
          >{\raggedleft}p{1.8cm}
          >{\raggedleft}p{1.8cm}
        | >{\raggedleft}p{1.8cm}
          >{\raggedleft}p{1.8cm}
          r|
    }
    \hline
    System & $(i,j)$ 
    & $t_{ij}$ (HSE) & $U_{ij}$ (HSE) & $J_{ij}$ (HSE)
    & $t_{ij}$ (PBE) & $U_{ij}$ (PBE) & $~~~J_{ij}$ (PBE) \\
    \hline
    ZrV in w-AlN
    & (1,1) & 11.0490 & 2.6258 &        & 11.3439 & 2.0787 & \\
    & (2,2) & 11.1031 & 2.5854 &        & 11.3867 & 2.0493 & \\
    & (3,3) & 11.0487 & 2.6258 &        & 11.3459 & 2.0787 & \\
    & (1,2) & -0.9938 & 1.8957 & 0.2009 & -0.7930 & 1.4343 & 0.1754 \\
    & (1,3) &  1.0312 & 1.9006 & 0.1945 &  0.8253 & 1.4382 & 0.1695 \\
    & (2,3) &  0.9940 & 1.8957 & 0.2009 &  0.7920 & 1.4344 & 0.1754 \\
    \hline
    HfV in w-AlN
    & (1,1) & 11.1122 & 2.5571 &        & 11.4455 & 2.0622 &  \\
    & (2,2) & 11.1748 & 2.5520 &        & 11.4966 & 2.0282 &  \\
    & (3,3) & 11.1134 & 2.5571 &        & 11.4469 & 2.0622 &  \\
    & (1,2) & -1.0240 & 1.9095 & 0.1912 & -0.8384 & 1.4148 & 0.1796\\
    & (1,3) &  1.0660 & 1.9088 & 0.1893 &  0.8769 & 1.4299 & 0.1742 \\
    & (2,3) &  1.0234 & 1.9096 & 0.1912 &  0.8378 & 1.4149 & 0.1797 \\
    \hline
    TiV in w-AlN
    & (1,1) & 10.8699 & 2.6249 &        & 11.2003 & 2.0793 &  \\
    & (2,2) & 10.9352 & 2.6285 &        & 11.2463 & 2.0843 &  \\
    & (3,3) & 10.8700 & 2.6249 &        & 11.2005 & 2.0793 &  \\
    & (1,2) & -0.9840 & 2.0563 & 0.2244 & -0.7986 & 1.5643 & 0.1998 \\
    & (1,3) &  1.0236 & 2.0563 & 0.2241 & 0.8308 & 1.5651 & 0.1994 \\
    & (2,3) &  0.9840 & 2.0563 & 0.2244 & 0.7985 & 1.5643 & 0.1998 \\
    \hline
    NV in diamond 
    & (1,1), (2,2), (3,3) & 15.3843 & 3.1241 & & 15.6472 & 2.6656 &  \\
    & (1,2) & -0.7049 & 1.0826 & 0.1416 & -0.5176 & 0.8431& 0.1635  \\
    & (1,3), (2,3) &  0.7049 & 1.0826 & 0.1416 & 0.5176 & 0.8431 & 0.1635  \\
    \hline
    %\end{tabularx}
    \end{tabular}
    \label{table:model_parameters}
\end{table*}

We summarized the model parameters of the low-energy effective Hamiltonian for $(\mathrm{Zr_{Al}V_{N}})^{0}$, $(\mathrm{Hf_{Al}V_{N}})^{0}$, $(\mathrm{Ti_{Al}V_{N}})^{0}$ complexes in w-AlN, and the NV$^-$-center in diamond in Table \ref{table:model_parameters}.
The model parameters were obtained using \texttt{RESPACK} \cite{Nakamura2021CPC}.
The KS wave functions as an input of \texttt{RESPACK} were obtained by spin-restricted DFT calculations with the PBE and HSE functional starting from the optimized structure.
We used cutoff energies for the dielectric functions of 3.0 Ry and 4.5 Ry for the NV center in diamond and complex defects in w-AlN, respectively.
The energy bands are considered for 750 and 1300 bands for the NV$^{-}$-center in diamond and complex defects in w-AlN, respectively.

\subsection{Specific expression of Hamiltonian}
The generator of the CRTE used in the main text is divided into a diagonal part $\Lambda$ and a non-diagonal part $V$:
$
    \mathcal{H} \otimes Z
    =
    \Lambda + V.
$
The diagonal part $\Lambda$ is divided as 
$
    \Lambda = \Lambda_1 + \Lambda_2, 
$
where
\begin{gather}
    \Lambda_1
    =
    h_{IIZZ} Z_{1}  Z_{0}
    +
    h_{IZIZ} Z_{2}  Z_{0}
    +
    h_{ZIZZ} Z_{3}  Z_{1}  Z_{0}
\end{gather}
and 
\begin{gather}
    \Lambda_2
    =
    h_{ZZIZ} Z_{3}  Z_{2}  Z_{0}
    +
    h_{ZIIZ} Z_{3}  Z_{0}
    \notag \\
    +
    h_{IZZZ} Z_{2}  Z_{1}  Z_{0}
    +
    h_{ZZZZ} Z_{3}  Z_{2}  Z_{1}  Z_{0} .
\end{gather}
The non-diagonal part $V$ is divided as 
$
    V = V_1 + V_2, 
$
where
\begin{gather}
    V_1
    =
    h_{XXXZ} X_{3} X_{2} X_{1} Z_{0}
    +
    h_{ZZXZ} Z_{3} Z_{2} X_{1} Z_{0}
    \nonumber \\
    +
    h_{ZXZZ}  Z_{3} X_{2} Z_{1} Z_{0}
    +
    h_{XZZZ}  X_{3} Z_{2} Z_{1} Z_{0}
    \nonumber \\
    +
    h_{IXXZ}  X_{2} X_{1} Z_{0}
    +
    h_{ZXXZ}  Z_{3} X_{2} X_{1} Z_{0}
    \nonumber \\
    +
    h_{XZXZ}  X_{3} Z_{2} X_{1} Z_{0}
    +
    h_{XIXZ}  X_{3} X_{1} Z_{0}
    \nonumber \\
    +
    h_{XXZZ}  X_{3} X_{2} Z_{1} Z_{0}
    +
    h_{XXIZ}  X_{3} X_{2} Z_{0}
    +
    h_{IIXZ}  X_{1} Z_{0}
    \nonumber \\
    +
    h_{IXIZ}  X_{2} Z_{0}
    +
    h_{XIIZ}  X_{3} Z_{0}
    +
    h_{IXZZ} X_{2} Z_{1} Z_{0}
    \nonumber \\
    +
    h_{ZXIZ} Z_{3} X_{2} Z_{0}
\end{gather}
and 
\begin{gather}
    V_2
    =
    h_{YYXZ} Y_{3} Y_{2} X_{1} Z_{0}
    +
    h_{YXYZ} Y_{3} X_{2} Y_{1} Z_{0}
    \nonumber \\
    +
    h_{XYYZ} X_{3} Y_{2} Y_{1} Z_{0}
    +
    h_{IZXZ} Z_{2} X_{1} Z_{0}
    +
    h_{YZYZ} Y_{3} Z_{2} Y_{1} Z_{0}
    \nonumber \\
    +
    h_{YIYZ} Y_{3} Y_{1} Z_{0}
    +
    h_{YYZZ} Y_{3} Y_{2} Z_{1} Z_{0}
    +
    h_{YYIZ} Y_{3} Y_{2} Z_{0}
    \nonumber \\
    +
    h_{ZIXZ} Z_{3} X_{1} Z_{0}
    +
    h_{XIZZ} X_{3} Z_{1} Z_{0}
    +
    h_{XZIZ} X_{3} Z_{2} Z_{0}.
\end{gather}
The Pauli operators $\{P_{\ell}\}$ are the same, with only the coefficients $\{h_{\ell}\}$ differing depending on the system.
The above generator is obtained by the function \texttt{Z2Symmetries} in \texttt{qiskit} v1.1.0 \cite{Qiskit}, which is applied to the low-energy effective Hamiltonian specified by the parameters in Table \ref{table:model_parameters} after the Pauli transformation.
% The coefficients of the generator $\{h_{\ell}\}$ are summarized in XX.

\section{Numerical results of other vacancies}
\label{sec:numerical_results_of_other_vacancies}
\subsection{FCI calculations}

We calculated other complex defects in $w$-AlN that are promising candidates for spin qubits.
We selected the previously reported $\mathrm{(Zr_{Al}V_{N})^{0}}$ \cite{Varley2016PRB, Seo2017PRM}, $\mathrm{(Ti_{Al}V_{N})^{0}}$ \cite{Varley2016PRB}, and $\mathrm{(Hf_{Al}V_{N})^{0}}$ \cite{Seo2017PRM} complexes in $w$-AlN. 
We used the same calculation settings for the DFT and RESPACK calculations, as in Sec. \ref{sec:method_classical} in main text.

\begin{figure*}
    \centering
    \includegraphics[width=0.95 \textwidth]{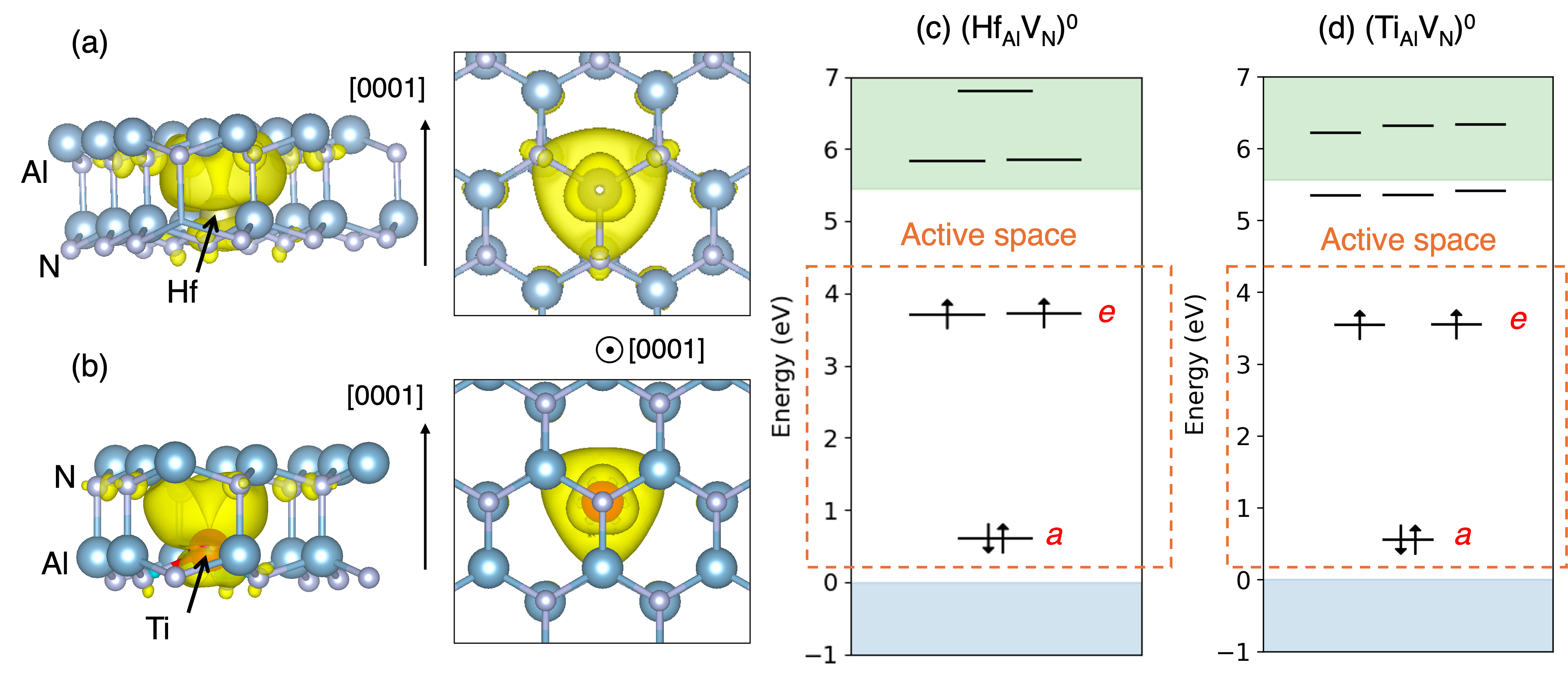} 
\caption{
Spin densities of (a) $\mathrm{(Hf_{Al}V_{N})^{0}}$ and (b) $\mathrm{(Ti_{Al}V_{N})^{0}}$ in w-AlN obtained by spin-unrestricted DFT calculations.
Isosurfaces are displayed at 10\% and 2.5\% of the maximum values for $\mathrm{(Hf_{Al}V_{N})^{0}}$ and (b) $\mathrm{(Ti_{Al}V_{N})^{0}}$, respectively, in w-AlN.
Kohn-Sham defect levels of (c) $\mathrm{(Hf_{Al}V_{N})^{0}}$ and (d) $\mathrm{(Ti_{Al}V_{N})^{0}}$ in-wAlN obtained by spin restricted DFT calculation. 
Orange dashed box denotes an energy region to take into account for constructing low-energy effective Hamiltonian. 
}
\label{fig:dft_other_aln}
\end{figure*}

The spin densities of $\mathrm{(Hf_{Al}V_{N})^{0}}$ and $\mathrm{(Ti_{Al}V_{N})^{0}}$ calculated from spin-polarized DFT calculations are presented in Fig. \ref{fig:dft_other_aln}(a) and (b), respectively.
The KS defect states obtained from spin-restricted DFT calculations are shown in Fig. \ref{fig:dft_other_aln}(c) and (d), respectively.
The results of the FCI calculations for the low-energy effective model constructed for the active space, framed by the orange dotted lines in Figures \ref{fig:dft_other_aln} (c) and (d), are summarized in Table \ref{table:excitation_energy_all_aln}.

\begin{table*}[ht]
    \centering
    \renewcommand{\arraystretch}{1.2} % Adjust row height
    \setlength{\tabcolsep}{8pt} % Adjust column spacing
    \newcolumntype{Y}{>{\centering\arraybackslash}X}
    \caption{Energies (eV) calculated by FCI calculation for the complex defects in w-AlN. In cases where the FCI energy is slightly split, the energy of one of the two is given in brackets.}
    \begin{tabular}{|m{1.5cm} m{1.5cm}|m{1.7cm} m{1.7cm} m{1.7cm} m{1.7cm} m{1.7cm}|}
        \hline
        vacancy & functional 
        & ${}^3A_2 \to {}^3E$  
        & ${}^3A_2 \to {}^1A_1$
        & ${}^3A_2 \to {}^1E$  
        & ${}^1E   \to {}^1A_1$
        & ${}^1A_1 \to {}^3E$ 
        \\  \hline
        $\mathrm{(Zr_{Al}V_{N})^{0}}$
        & PBE & 2.39 (2.43) & 0.86 & 0.43 (0.44) & 0.43 (0.42) & 1.54 (1.57)
        \\ 
        & HSE & 3.00 (3.04) & 0.98 & 0.49 (0.5) & 0.48 (0.49) & 2.02 (2.06)
        \\ \hline
        $\mathrm{(Ti_{Al}V_{N})^{0}}$
        & PBE & 2.53 (2.58) & 0.87 & 0.43 (0.44) & 0.43 (0.44) & 1.66 (1.71)
        \\ 
        & HSE & 3.11 (3.12) & 0.91 & 0.46 (0.47) & 0.45 (0.45) & 2.19 (2.21) 
        \\ \hline
        $\mathrm{(Hf_{Al}V_{N})^{0}}$
        & PBE & 2.42 (2.43) & 0.85 & 0.44 (0.44) & 0.41 (0.42) & 1.57 (1.58) 
        \\ 
        & HSE & 2.99 (3.00) & 0.95 & 0.49 (0.49) & 0.46 (0.47) & 2.04 (2.04)
        \\ \hline
    \end{tabular}
    \label{table:excitation_energy_all_aln}
\end{table*}

\subsection{Quantum computation of ground state of the singlet state}
\subsubsection{Trunction error on qubit reduction using $\mathbb{Z}_2$ symmetry}
As in the case of the $\mathrm{(Zr_{Al}V_{N})^{0}}$ complex defects in w-AlN described in the main text, the truncation values of the Hamiltonians for $\mathrm{(Hf_{Al}V_{N})^{0}}$ and $\mathrm{(Ti_{Al}V_{N})^{0}}$ defects are small, at 0.01 eV and 0.005 eV, respectively.
As a result, we did not observe any effect of truncation on the FCI energy at the second decimal point.
On the other hand, the value of the truncation for NV centers in diamonds is somewhat large at 0.07 eV, and the FCI energy of the ${}^1 E$ state increases slightly from 0.64 eV to 0.7 eV.
In this truncation, the many-body wave function of ${}^1 E$ state for NV-center diamond, presented in Table \ref{table:wave_function_nv} in main text, is changed as 
\begin{gather}
    0.69 \big(
        |e_{y\uparrow} e_{x\downarrow} a_{1\uparrow} a_{1 \downarrow}\rangle
        -
        |e_{y\downarrow} e_{x\uparrow} a_{1\uparrow} a_{1 \downarrow}\rangle
    \big)
    \notag \\
    +
    0.17 \big(
        |e_{y\uparrow} e_{y\downarrow} e_{x\downarrow} a_{1 \uparrow}\rangle
        -
        |e_{y\uparrow} e_{y\downarrow} e_{x\uparrow} a_{1 \downarrow}\rangle
    \big)
\end{gather}
and 
\begin{gather}
    0.69 \big(
        |e_{x\uparrow} e_{x\downarrow} a_{1\uparrow} a_{1 \downarrow}\rangle
        -
        |e_{y\uparrow} e_{y\downarrow} a_{1\uparrow} a_{1 \downarrow}\rangle
    \big)
    \notag \\
    +
    0.17\big(
        |e_{y\downarrow} e_{x\uparrow} e_{x\downarrow} a_{1 \uparrow}\rangle
        -
        |e_{y\uparrow} e_{x\uparrow} e_{x\downarrow} a_{1 \downarrow}\rangle
    \big) .
\end{gather}

\subsubsection{Ground-state calculation using emulator of H1-1}
\begin{figure*}
    \centering
    \includegraphics[width=1.0 \textwidth]{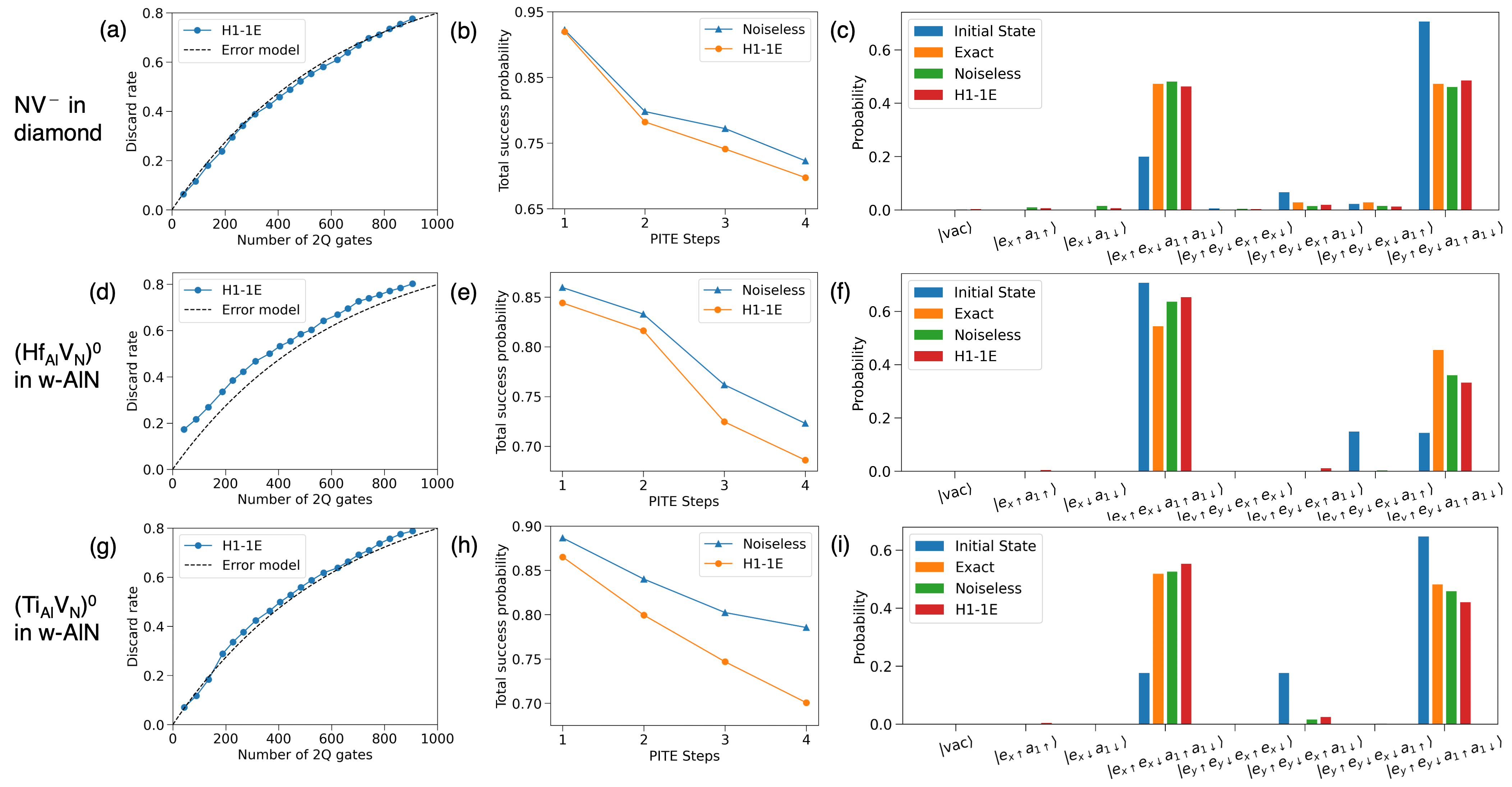} 
\caption{
Quantum computation of the ground state of the spin singlet state for (a-c) NV${}^{-}$-center in diamond, (d-f) $(\mathrm{Hf_{Al}V_{N}})^{0}$ complex in w-AlN, and (g-i) $(\mathrm{Ti_{Al}V_{N}})^{0}$ complex in w-AlN.
(a, d, g) Discard rates of the calculated results owing to errors detected by QED as a function of the number of two-qubit gates. The dashed line denotes noise model, as shown in Eq. (\ref{eq:noise_fit}) with $p_2 = 1.6 \times 10^{-3}$. 
(b, e, h) Total success probability of the PITE according to the PITE steps. 
(c, f, i) Histogram of measured computational basis. 
Noiseless denotes the noiseless simulation of PITE, and H1-1E denotes the emulator of H1-1 quantum computer.
All calculations used $N_{\mathrm{shot}} = 2000$. 
}
\label{fig:result_other_gs}
\end{figure*}

We present the computed results of the ground state of the singlet state on the emulator of H1-1 for the NV-center diamond in Fig. \ref{fig:result_other_gs}(a-c), HfV complex in w-AlN in Fig. \ref{fig:result_other_gs}(d-f), TiV complex in w-AlN in Fig. \ref{fig:result_other_gs}(g-i).
The calculation conditions are the same as those described in Sec. \ref{sec:results_quantum} of the main text, except that $N_{\mathrm{shots}} = 2000$.  
The values of the calculation results in Figure \ref{fig:result_other_gs} are summarized in Table \ref{table:result_other_gs}.

\begin{table}[ht]
    \centering
    \renewcommand{\arraystretch}{1.2} % Adjust row height
    \setlength{\tabcolsep}{1pt} % Adjust column spacing
    \newcolumntype{Y}{>{\centering\arraybackslash}p{1.5cm}}
    \caption{Computed results for the ground state of the spin singlet state for NV center in diamond, TiV in w-AlN, and HfV in w-AlN on the emulator of H1-1 (H1-1E). Total success probability $\mathbb{P}_{\mathrm{tot}}$, classical fidelity $F$, and discard rate $d$ are summarized.}
    %\begin{tabular}{|Y|YY|YY|Y|}
    \begin{tabular}{
        |m{1.5cm}|
        m{1.3cm}
        m{1.3cm}|
        m{1.3cm}
        m{1.3cm}|
        m{1.3cm}|
    }
        \hline
        % \rowcolor{gray}
        & \multicolumn{2}{|c|}{$\mathbb{P}_{\mathrm{tot}}$}   & \multicolumn{2}{|c|}{$F$}  & $d$
        \\ \cline{2-6}
        System & Noiseless & H1-1E & Noiselss & H1-1E & H1-1E
        \\  \hline
        $\mathrm{NV^{-}}$
        & 0.72 & 0.70 & 0.97 & 0.98 & 0.78
        \\  
        $\mathrm{(Hf_{Al}V_{N})^{0}}$
        & 0.79 & 0.70 & 0.98 & 0.97 & 0.79 
        \\ 
        $\mathrm{(Ti_{Al}V_{N})^{0}}$
        & 0.72 & 0.69 & 0.99 & 0.98 & 0.80
        \\
        \hline
    \end{tabular}
    \label{table:result_other_gs}
\end{table}

\bibliography{ref}

\end{document}